\documentclass[journal]{IEEEtran}
\usepackage{xcolor,soul,framed} %,caption
\usepackage{cite}
\usepackage{amsmath,amssymb,amsfonts}
\usepackage{gensymb}
\usepackage{graphicx}
\usepackage{textcomp}
\usepackage{dcolumn}% Align table columns on decimal point
\usepackage{bm}% bold math
\usepackage{subfigure} %Para poder hacer subplots
\usepackage{pict2e}
\usepackage{epstopdf}
\usepackage{arydshln}
\usepackage{comment}

\begin{document}

\title{A Dual-Mode FM/AM Modulator Based on a Time-Varying Inverting Integrator}

\author{Azalía G. Gil, Alfonso T. Muriel-Barrado, Mario Pérez-Escribano, \\ Carlos Molero, Antonio Alex-Amor  % <-this % stops a space
\thanks{This work was supported in part by the BBVA Foundation's Leonardo Grant for Scientific Research and Cultural Creation 2025, in part by the Grant PID2024-155167OA-I00 funded by MICIU/AEI/10.13039/501100011033/FEDER, UE, and in part by Consejería de Universidad, Investigación e Innovación of Junta de Andalucía through grant EMERGIA 23-00235. The BBVA Foundation is not responsible for the opinions, comments, and content included in the project and/or the results derived from it, which are the sole and absolute responsibility of their authors.}
\thanks{A.G. Gil, A.T. Muriel-Barrado, and A. Alex-Amor are with the Department of Electronic and Communication Technology, RFCAS Research Group,  Universidad Autónoma de Madrid, 28049 Madrid, Spain.}
\thanks{M. Pérez-Escribano is with Dpto. de Ingeniería de Comunicaciones, Telecommunication Research Institute (TELMA),
Universidad de Málaga, E.T.S. Ingeniería de Telecomunicación, 29010 Málaga, Spain.}
\thanks{C. Molero is with the Department of Electronics and
Electromagnetism, Physics Faculty, Universidad de Sevilla, 
Avenida de la Reina Mercedes S/N, 41012, Sevilla, Spain.}
}

\markboth{}%
{}

\maketitle

\newcommand*{\bigs}[1]{\vcenter{\hbox{\scalebox{2}[8.2]{\ensuremath#1}}}}

\newcommand*{\bigstwo}[1]{\vcenter{\hbox{ \scalebox{1}[4.4]{\ensuremath#1}}}}

\begin{abstract}
This paper presents the analysis, design, fabrication, and experimental validation of a dual-mode frequency/amplitude modulator  based on a time-modulated varactor diode. By exploiting the varactor as a time-varying capacitor in combination with an operational amplifier configured as an inverting integrator and a passband filter, the proposed circuit generates frequency-modulated (FM)  signals in an efficient manner. Amplitude-modulated (AM) signals can also be obtained with a simple modification. The implementation, realized in microstrip technology, leverages the unique properties of time-modulated electronic components, particularly their inherent frequency-mixing capability. Analytical expressions are derived to predict the characteristics of the generated waveforms, and their accuracy is verified through numerical simulations performed in Keysight ADS. A microstrip PCB prototype is then fabricated and experimentally characterized. The measured results show excellent agreement with both the theoretical predictions and the numerical simulations. The proposed approach demonstrates the potential of time-varying capacitors as an attractive alternative to conventional FM techniques for telecommunications and radar applications.
\end{abstract}


\IEEEpeerreviewmaketitle

%%%%%%%%%%%%%%%%%%%%%%%%%%%%%%%%%%%%%%%%%%%%%%%%
\section{Introduction}
%%%%%%%%%%%%%%%%%%%%%%%%%%%%%%%%%%%%%%%%%%%%%%%%
Telecommunications represents one of the major advances that have allowed people to interconnect and improve their lives. The International Telecommunication Union (ITU) says that in 2025 the online population has reached 6 billion people. In addition, 5G networks cover more than 55\% of the world's population, which implies a massive connectivity demand. This situation, which has been developing since 2021, forces the development of more efficient and low-cost solutions towards next-generation 6G networks so that this constant saturation can be managed \cite{international_telecommunication_union_measuring_2025, rasilainen_hardware_2023, chamola_fpga_2020}.

Generally, the hardware of telecommunications systems is based on stationary materials and electronic components, i.e., time-invariant elements that can change their electromagnetic properties in space but not in time \cite{kong_electromagnetic_2008, pozar_microwave_2021, balanis_balanis_2024}.  Time-invariant devices, while  simpler to design and implement, are restricted by passivity conditions that limit their performance, such as Lorentz reciprocity, or Bode-Fano and Rozanov bounds on available bandwidth in matching and absorption \cite{balanis_balanis_2024, fano_theoretical_1950, rozanov_ultimate_2000}.

Motivated by these fundamental limitations, considerable research efforts have recently been devoted to exploiting the unique capabilities of time-varying devices to develop next-generation communication and radar systems in both the radiofrequency (RF) and optical regimes \cite{galiffi_photonics_2022, caloz_spacetime_2020, caloz_spacetime_2020-1, taravati_microwave_2022, engheta_four-dimensional_2023, AlexAmorOptics26}. These capabilities, which mainly include frequency conversion, magnet-free nonreciprocity, parametric amplification, and improved bandwidth-size ratio \cite{Fan2009, wu_serrodyne_2020, prudencio_synthetic_2023, moreno-rodriguez_space-time_2024, ramaccia_nonreciprocity_2018, yuh_sun_direct_1981, pendry_gain_2021, lyubarov_amplified_2022, hayran_using_2022, shlivinski_beyond_2018, li_temporal_2021},  have led to the implementation of modern time-modulated devices acting as nonreciprocal filters, RF isolators, space-time-coded metasurfaces, and nonreciprocal antennas, to name a few \cite{wu_isolating_2019, estep_magnetic-free_2014, zhang_space-time-coding_2018, shihan_qin_nonreciprocal_2014, devitt_mems_2026, zang_nonreciprocal_2019}.

Among the functionalities enabled by time-varying devices, frequency conversion is particularly relevant to telecommunication systems, where spectral translation and modulation are fundamental operations. In this context, frequency modulation (FM) remains one of the most widely used analog modulation techniques owing to its robustness against noise, constant-envelope operation, and resilience to interference \cite{jiang_semantic_2025, maral_satellite_2020, kotelnikov_theory_1968}. Consequently, FM continues to play a key role in a wide range of applications, including frequency-modulated continuous-wave (FMCW) radar systems, satellite communications, and commercial radio broadcasting, as well as telemetry and sensing \cite{scheiblhofer_high-speed_2007, mitomo_77_2010, molchanov_short-range_2015, peng_sinusoidal_2014, balcells_emi_2005, GarciaArmadaFM2023, WangFM2024, LiFM2025}. Amplitude modulation (AM) represents another fundamental class of analog modulation techniques that, despite being one of the earliest modulation schemes, remains relevant in many scenarios, including broadcasting, optical and wireless communication systems, and signal processing, due to its simplicity compared to FM \cite{haykin2001communication}.

Typically, FM is achieved using well-established circuits such as voltage-controlled oscillators (VCOs) \cite{choi_24_2003, brennan_determination_2011, tsitouras_ultra_2009, riaz_modulation_2026}. In these circuits, the oscillation frequency is directly controlled by an input voltage, so that variations in the control signal are translated into corresponding changes in the instantaneous output frequency. Another conventional architecture is the Armstrong indirect frequency modulator, in which a narrowband FM (NBFM) signal is initially obtained through a phase-modulation stage. Then a wideband FM (WBFM) signal is achieved through a chain of frequency multipliers. These frequency multipliers use nonlinear devices to generate harmonics of the input signal, combined with a bandpass filter to select the desired harmonic, so that both the carrier frequency and the frequency deviation are scaled by the same factor \cite{armstrong_evolution_1940, lewis_shoulders_2021, schwartz_armstrongs_2009, armstrong_method_1936}. In turn, AM generally enables simpler circuit implementations than FM, with conventional modulators typically based on diode bridges or transistor circuits that exploit nonlinear device characteristics to control the carrier amplitude \cite{lathi2009modern, kennedy2008electronic}.

In digital hardware, some software-defined radio (SDR) architectures implemented on field-programmable gate arrays (FPGAs) have shown the viability of high-performance FM and AM modulations. These modulators and demodulators stand out because of the use of direct digital frequency synthesizers (DDFS) optimized by quarter-wave symmetry techniques \cite{Vankka2001DDS}, achieving a spurious free dynamic range (SFDR) of greater than 64~dB. In this way, efficient hardware implementations for real-time FM and AM signal processing can be realized \cite{hatai_new_2011}.

In this work, we present a novel implementation of analog FM and AM modulators based on time-modulated circuit implementations. For this purpose, we exploit the frequency conversion properties of modulated varactors acting as time-varying capacitors. Combined with an operational amplifier configured as a modified inverting integrator and a lumped-element bandpass filter, the proposed architecture directly generates frequency-modulated signals with dynamically tunable characteristics. This work demonstrates that time-varying reactive elements provide an alternative physical mechanism to modulate signals, extending the application of time-varying circuits beyond the nonreciprocal and frequency-conversion devices commonly reported in the literature.

The work is organized as follows. Section~II presents the analysis of the main time-varying circuit that leads to FM modulation. Analytical formulas are provided to estimate the output FM signal, and the theory is compared to numerical simulations in software ADS. Section~III introduces an alternative circuit topology that simplifies the practical implementation and fabrication of the proposed FM modulator. It is shown how the alternative circuit can also produce amplitude-modulated (AM) signals. Section~IV describes the fabrication and experimental characterization of the prototype, with the measured results compared against the theoretical predictions and simulation results. Finally, Section~V summarizes the main conclusions of this work.

%%%%%%%%%%%%%%%%%%%%%%%%%%%%%%%%%%%%%%%%%%%%%%%%
\section{FM Generation via a Time-varying Circuit}
%%%%%%%%%%%%%%%%%%%%%%%%%%%%%%%%%%%%%%%%%%%%%%%%

This section details a first time-varying circuit topology that leads to FM modulation. The circuit, formed by a resistor $R$, a time-varying capacitor $C(t)$, and a bandpass filter, is illustrated in Fig.~\ref{fig:circuito_ideal_1}. To meet our goal, we will consider that the time-varying capacitor is sinusoidally-modulated according to
\begin{equation} \label{eq:ct}
    C(t) = C_0 + \Delta C \cos(\omega_c t),
\end{equation}
where $C_0$ is the nominal capacitance, $\Delta C$ is the amplitude factor of the time modulation, and $\omega_c$ is the carrier frequency.

%-----------------------------
\subsection{Theory}

\begin{figure}[!t]
    \centering
    \includegraphics[width=1\linewidth]{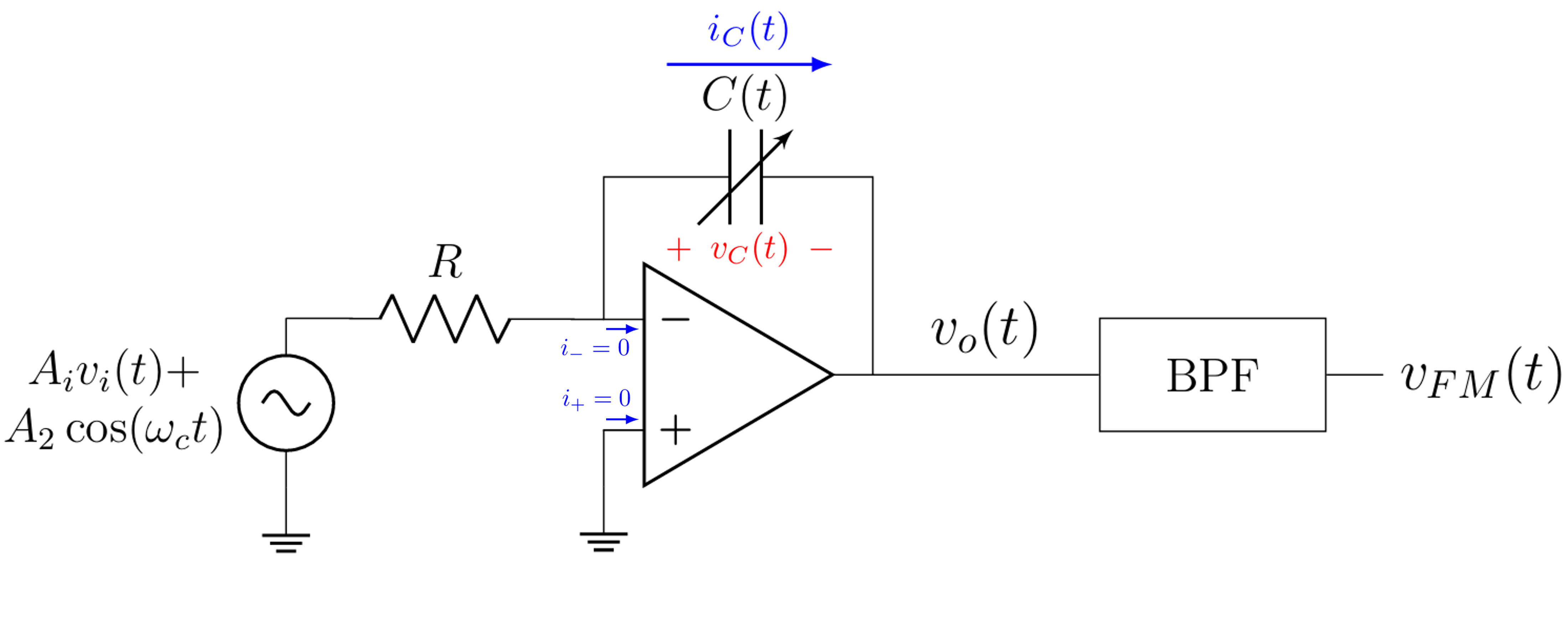}
    \caption{Time-varying circuit that produces an output FM waveform.}
    \label{fig:circuito_ideal_1}
\end{figure}

Before proceeding to the analysis of the circuit in Fig.~\ref{fig:circuito_ideal_1}, we should note that a generic FM signal has the form described in \cite[Ch.~3]{proakis_communication_2002}
\begin{equation}
    s_{FM}(t) = A_{FM} \sin\left(\omega_ct + k_{FM}\int^t_0 v_i(\tau)d\tau\right),
\end{equation}
where $A_{FM}$ is its peak amplitude, $k_{FM}$ is a frequency-sensitivity constant, and $v_i$ is the input message (baseband) signal.

The use of the trigonometric relation $\sin(a+b)=\sin(a)\cos(b)+\cos(a)\sin(b)$ in a scenario where $k_{FM} \ll 1$ (the cosine can be approximated by 1 and the sine by its argument) leads to
\begin{equation} \label{eq:sfm}
    s_{FM}(t) \approx A_{FM}\sin(\omega_ct) + A_{FM}k_{FM}\cos(\omega_ct)\int^t_0 v_i(\tau)d\tau.
\end{equation}

In the following, we will show that the circuit in Fig.~\ref{fig:circuito_ideal_1} is indeed capable of creating an output frequency-modulated signal. The time-varying capacitor is governed by the voltage-current relation $i_C(t) = \frac{d}{dt}[C(t)v_C(t)]$. Moreover, we consider the operational amplifier as an ideal element, thus: 1) its input impedance is infinity (no current enters its non-inverting ``$+$" and inverting ``$-$" terminals); 2) the virtual short-circuit technique can be applied as there is no positive feedback ($V_{+} = V_{-} = 0$). 

Therefore, if the input signal is $A_iv_i(t)+A_2\cos(\omega_ct)$, the current analysis of Fig.~\ref{fig:circuito_ideal_1} reveals that
\begin{equation}
    \frac{A_iv_i(t)+A_2\cos(\omega_ct)}{R} = -\frac{d}{dt}\Big[v_o(t)(C_0+\Delta C\cos(\omega_ct))\Big],
\end{equation}
where solving for the output voltage, $v_o(t)$, one obtains
\begin{equation}
    v_0(t) = -\frac{1}{R[C_0+\Delta C\cos(\omega_ct)]} \int^t_0\left(A_iv_i(\tau)+A_2\cos(\omega_c\tau)\right)d\tau.
\end{equation}

In the small-amplitude modulation regime, $\Delta C \ll C_0$, the output voltage can be simplified as
\begin{equation}
    v_o(t) \approx -\frac{C_0-\Delta C\cos(\omega_ct)}{RC_0^2} \int^t_0\left(A_iv_i(\tau)+A_2\cos(\omega_c\tau)\right)d\tau.
\end{equation}
Finally, expanding all the terms, the resultant voltage is
\begin{multline} \label{eq:vo_1}
    v_o(t) \approx -\frac{A_i}{RC_0}\int^t_0v_i(\tau)d\tau
    +\frac{\Delta CA_2}{2\omega_cRC_0^2}\sin(2\omega_ct) \\
    -\frac{A_2}{\omega_cRC_0}\sin(\omega_ct)
    +\frac{\Delta C A_i}{RC_0^2}\cos(\omega_ct)\int^t_0v_i(\tau)d\tau.
\end{multline}

A frequency analysis of Eq.~\eqref{eq:vo_1} reveals that the first right-hand term is a baseband signal ($v_i$ is the baseband signal and the integral acts as a low-pass filter), the second term is a tone centered at twice the carrier frequency, the third term is another tone centered just at the carrier frequency, and the fourth term is essentially a frequency-shifted replica of the first term, translated to the carrier frequency.

As seen, $v_0(t)$ in Eq.~\eqref{eq:vo_1} and the reduced FM signal $s_{FM}(t)$ in Eq.~\eqref{eq:sfm} exhibit similar spectral components, even though $v_0(t)$ cannot yet be considered a FM waveform. Specifically, the third and fourth terms in $v_0(t)$ are identical to  the corresponding components of $s_{FM}(t)$. Therefore, by appropriately filtering out the first and second terms of $v_0(t)$, the resulting output signal becomes equivalent to the desired FM waveform. To do so,  $v_0(t)$ must be injected through a bandpass filter (BPF) that preserves the spectral content around the carrier frequency $\omega_c$, while suppresses the baseband component and the harmonic term at $2\omega_c$. Thus, the modified integrator topology in Fig.~\ref{fig:circuito_ideal_1} with the time-varying capacitor and a bandpass filter connected at the output creates a FM signal, which can be described with the following expression
\begin{multline} \label{eq:vfm_1}
    v_{FM}(t) = \text{BPF}\{v_o(t)\} =
    -\frac{A_2}{\omega_cRC_0}\sin(\omega_ct) \\
    +\frac{\Delta C A_i}{RC_0^2}\cos(\omega_ct)\int^t_0v_i(\tau)d\tau.
\end{multline}
Comparing \eqref{eq:sfm} and \eqref{eq:vfm_1}, the FM amplitude and sensitivity parameters are given by
\begin{equation} \label{eq:afm_kfm_1}
    A_{FM} = -\frac{A_2}{\omega_cRC_0}, \quad k_{FM} = \frac{\Delta C A_i}{RC_0^2A_{FM}}.
\end{equation}

%-----------------------------
\subsection{Simulation}

Once the FM modulator has been analyzed, the next step is to check its correct performance in simulation. To achieve this, we perform some numerical simulations in commercial software ADS \cite{keysight_technologies_advanced_2025}. For the simulation, we select a baseband signal of the form $v_i(t) = \cos(2\pi f_{bb}t)$, with $f_{bb} = 30\,\text{MHz}$, and a carrier of $f_c = 100\,\text{MHz}$, in order to replicate the typical frequencies seen in commercial FM. The rest of the parameters are chosen as $C_0=1\,\text{pF}$ and $R=10\,\text{k}\Omega$. We first set \mbox{$\Delta C=0.1C_0$} and $k_{FM}=0.1\omega_c$, and later evaluate the cases with $\Delta C=0.3C_0$ and $k_{FM}=0.3\omega_c$.

\begin{figure}[!t]
  \centering
  \subfigure[]{
    \includegraphics[width=1\linewidth]{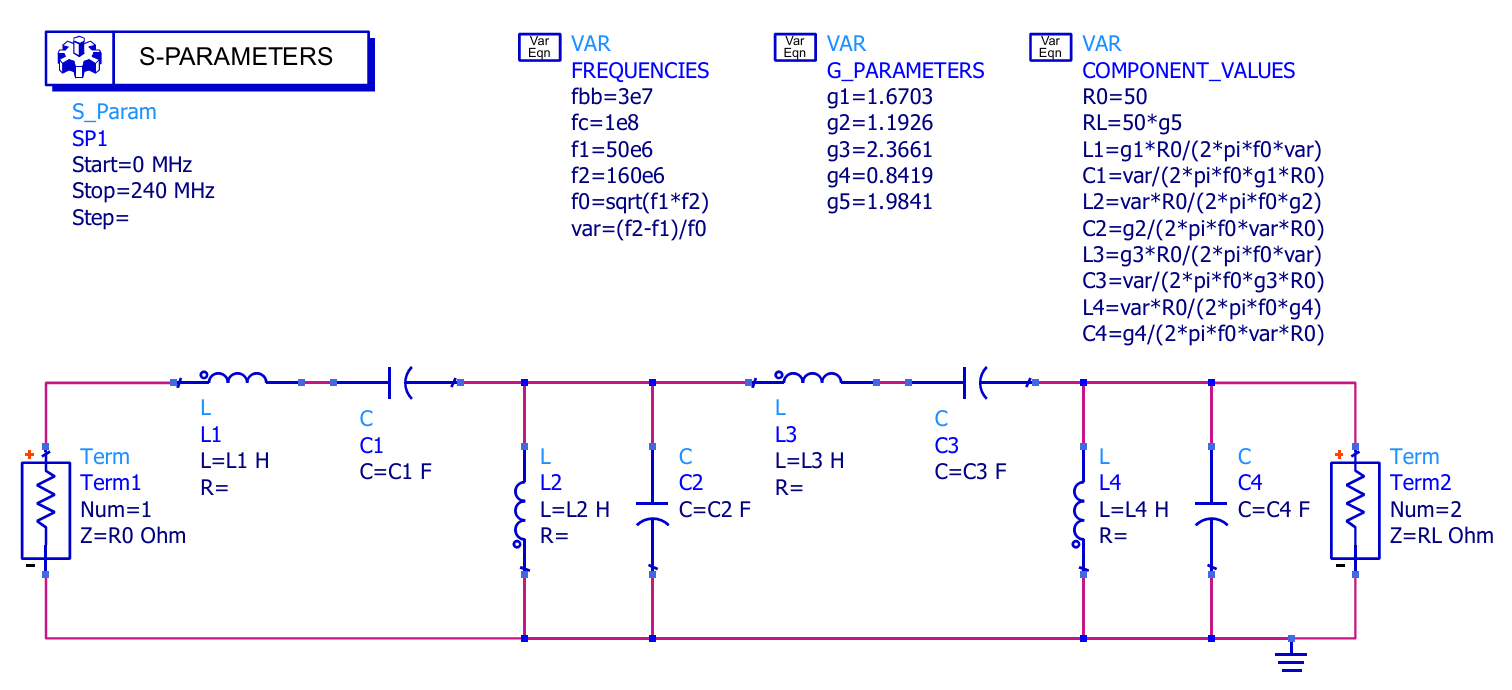}
    \label{fig:bpf_circuito}
  }
  
  \subfigure[]{
    \includegraphics[width=0.485\linewidth]{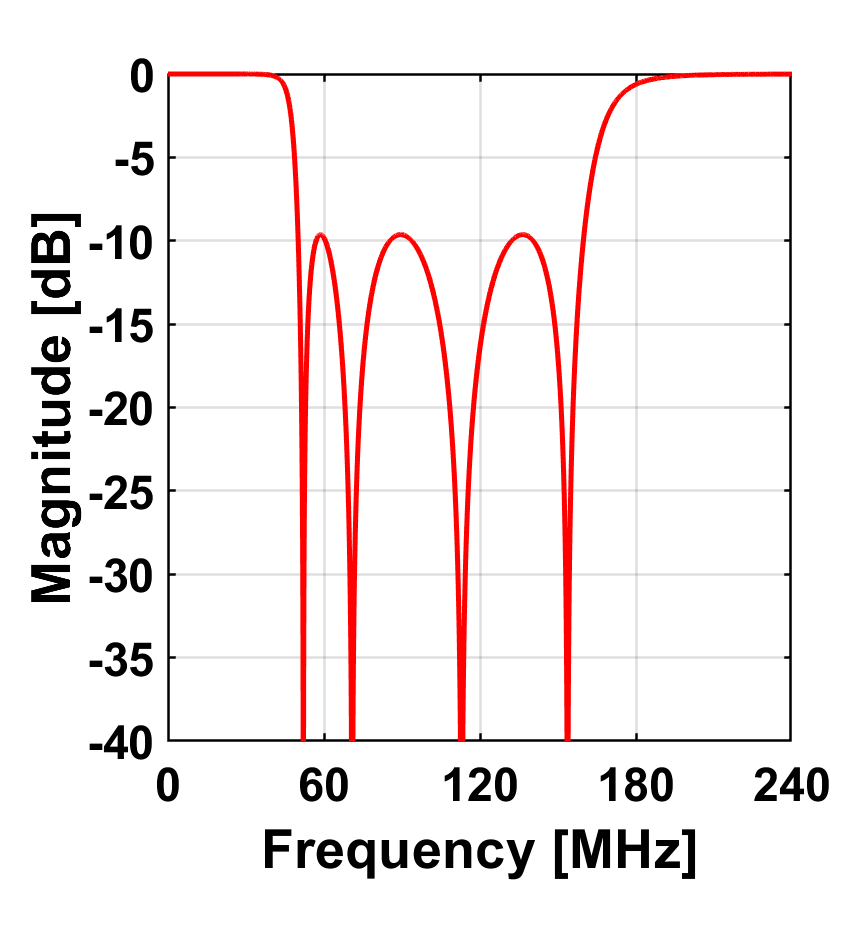}
    \label{fig:bpf_s11}
  }%
  \subfigure[]{
    \includegraphics[width=0.485\linewidth]{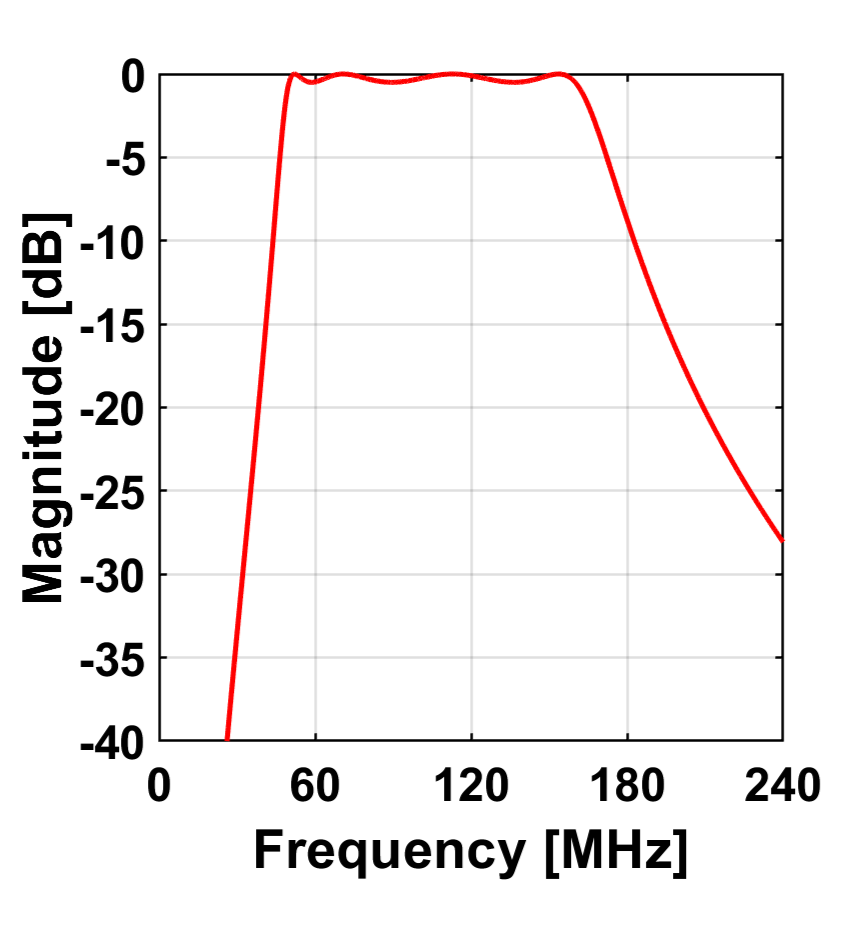}
    \label{fig:bpf_s21}
  }%
  
  \caption{Bandpass filter implemented in Fig. 1. \subref{fig:bpf_circuito} Schematic of the circuit in ADS. \subref{fig:bpf_s11} Simulated magnitude of the reflection coefficient $|S_{11}|$. \subref{fig:bpf_s21} Simulated magnitude of the transmission coefficient $|S_{21}|$.}
  \label{fig:bpf}
\end{figure}

We start by checking the correct operation of the bandpass filter. The type of filter that has been chosen for this assignment is a four-stage lumped-element Chebyshev filter with a 0.5 dB ripple following the procedure described in \cite[Ch.~8]{pozar_microwave_2021}. The characteristic impedance is set to $Z_0=50\,\Omega$, and the cutoff  frequencies are established at $f_1=50\,\text{MHz}$ and $f_2=160\,\text{MHz}$, yielding a center frequency of  $f_0 = \sqrt{f_1\cdot f_2} = 89.44\,\text{MHz}$. Consequently, the component values were obtained from the $g$ parameters and the required transformation. Finally, the filter topology and its  frequency response are shown in Fig.~\ref{fig:bpf}. As seen, the frequency response fits the requirements, as $30\,\text{MHz}$ and $200\,\text{MHz}$ are effectively filtered out. If needed, the number of stages can be improved to increase the rejection.

\begin{figure}[!t]
  \centering
  \subfigure[]{
    \includegraphics[width=1\linewidth]{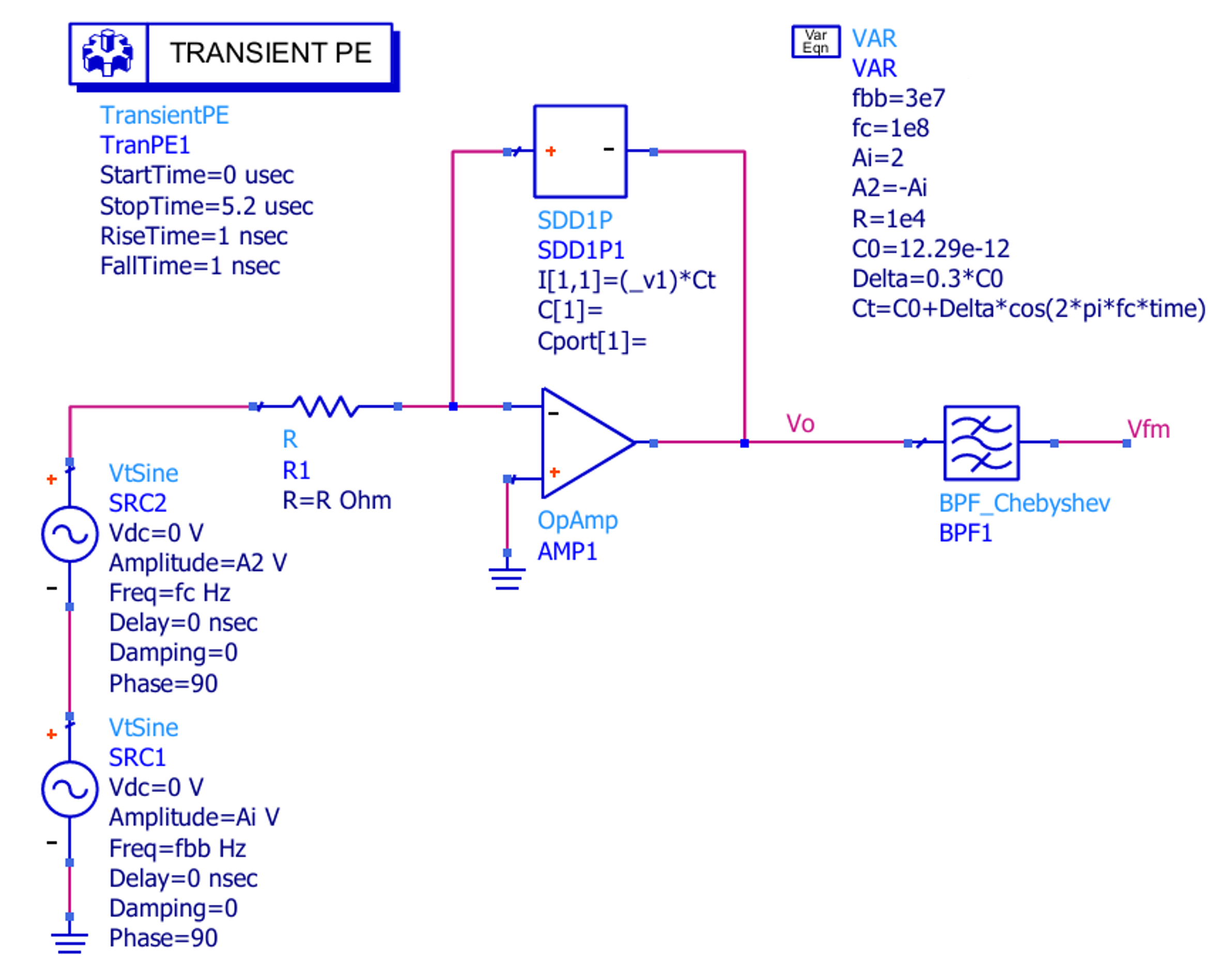}
    \label{fig:circuito_fm_ADS_1}
  }

    \subfigure[]{
    \includegraphics[width=1\linewidth]{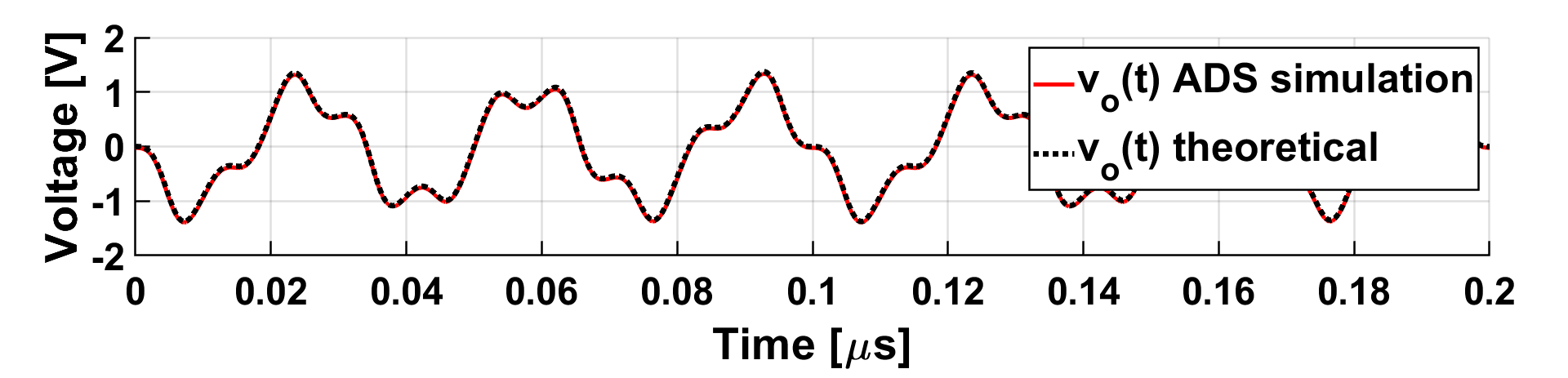}
    \label{fig:respuesta_vo10}
  }

    \subfigure[]{
    \includegraphics[width=1\linewidth]{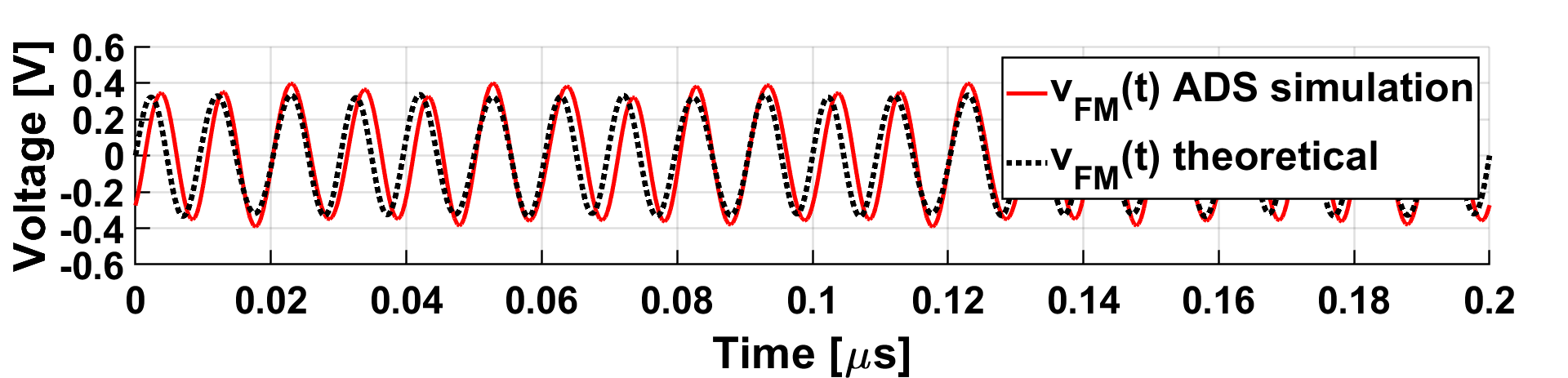}
    \label{fig:respuesta_vfm10}
  }
  
  \subfigure[]{
    \includegraphics[width=1\linewidth]{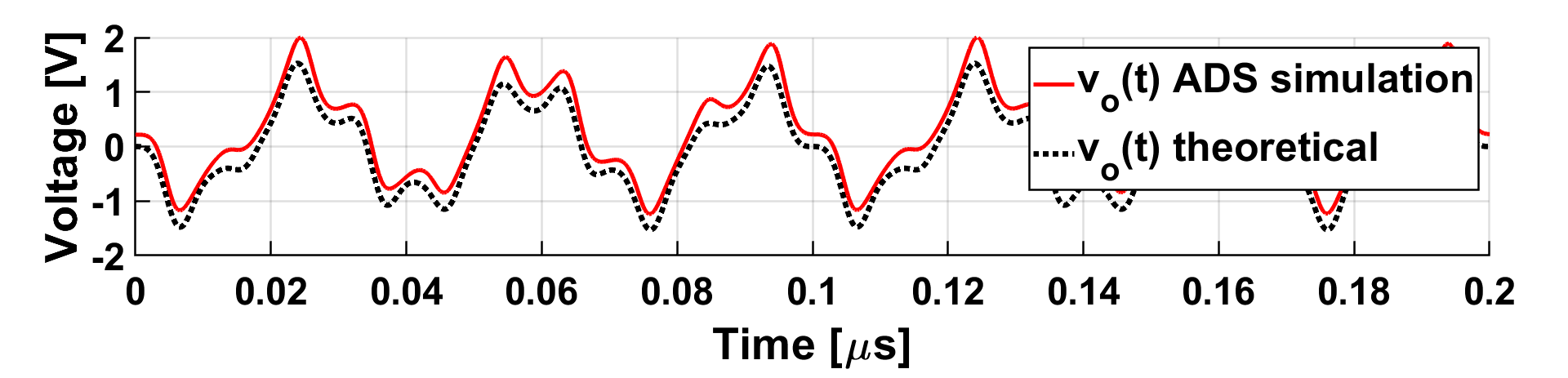}
    \label{fig:respuesta_vo30}
  }

    \subfigure[]{
    \includegraphics[width=1\linewidth]{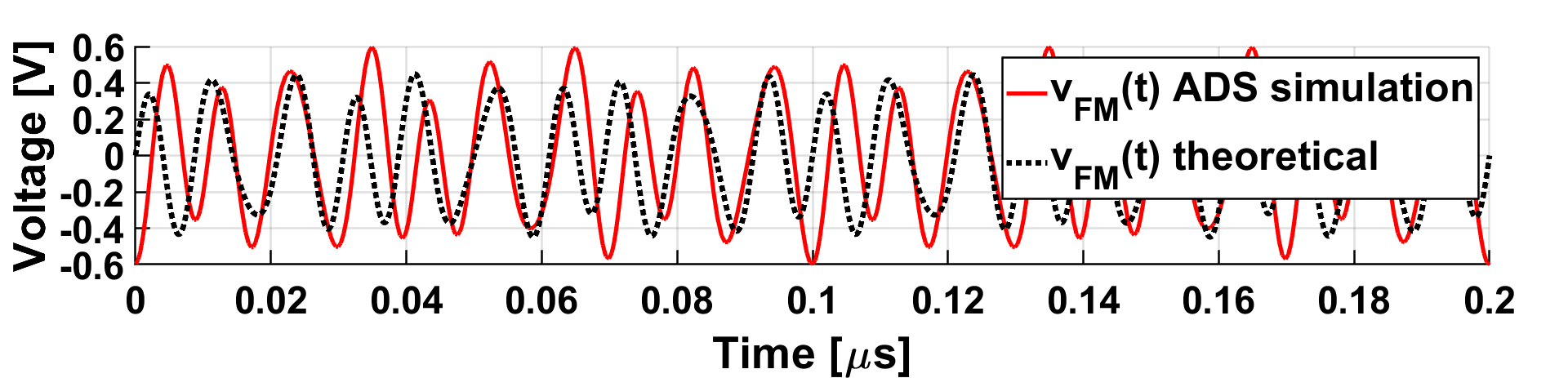}
    \label{fig:respuesta_vfm30}
  }
  
  \caption{First FM circuit with a time-varying capacitor. \subref{fig:circuito_fm_ADS_1} Schematic of the circuit in ADS. \subref{fig:respuesta_vo10} Time-domain response of the operational amplifier's output voltage for a modulation factor of 10\%. \subref{fig:respuesta_vfm10} Time-domain response of the FM output voltage for a modulation factor of 10\%. \subref{fig:respuesta_vo30} Time-domain response of the operational amplifier's output voltage for a modulation factor of 30\%. \subref{fig:respuesta_vfm30} Time-domain response of the FM output voltage for a modulation factor of 30\%.}
  \label{fig:fm_1}
\end{figure}

For the simulation of the FM modulator in Fig. 1, we employ the transient solver, as it is particularly suitable for analyzing the time-domain behavior of circuits containing time-varying elements. The complete circuit schematic in ADS is represented in Fig.~\ref{fig:fm_1}(a).  The time-modulated capacitor is implemented using the user-defined component “SDD1P”, available in ADS. This block enables the definition of a generic current–voltage relationship, allowing the behavior of the time-varying capacitor $C(t)$ to be accurately modeled through a user-defined equation. In this case, the component is configured with the parameter $I[p,w]$ where $p$ refers to the port number and $w$ refers to the weighting function \cite{agilent_technologies_advanced_2012}. Using the first port ($p=1$) and the first-order time derivative ($w=1$), the explicit equation is defined as ``$I[1,1]=(\_v1) \ast Ct$", where ``$Ct =C0+Delta \ast cos(wc \ast time)$" is given by Eq.~\eqref{eq:ct}. Moreover, the block ``BPF\_Chebyshev'' refers to the bandpass filter shown in Fig.~\ref{fig:bpf}(a).

The simulation results (red curves) are shown in Fig.~\mbox{\ref{fig:fm_1}\subref{fig:respuesta_vo10}-\subref{fig:respuesta_vfm30}} and then compared to the analytical expression (black dashed curve) given by Eq.~\eqref{eq:vfm_1}. Figures~\mbox{\ref{fig:fm_1}\subref{fig:respuesta_vo10}-\subref{fig:respuesta_vfm10}} present $v_0$ and the FM signal $v_{FM}$, respectively, for a modulation factor of 10\% ($\Delta C = 0.1C_0$). Figures~\mbox{\ref{fig:fm_1}\subref{fig:respuesta_vo30}-\subref{fig:respuesta_vfm30}} do the same for a modulation factor of 30\%. As seen, the agreement between theory and simulations is good, especially for small-amplitude modulations such as the 10\%, showing that the proposed circuit topology in Fig.~\ref{fig:circuito_ideal_1} indeed acts as a FM modulator.

It is also observed that the agreement between theory and simulation is better for $v_0$ than for $v_{FM}$. This discrepancy is attributed to the finite attenuation of the bandpass filter, which limits the suppression of the undesired spectral components. Increasing the number of filter stages is expected to further improve the rejection of these components and, consequently, enhance the overall performance of the proposed system.

Notably, the discrepancy between the expected signal and the simulated one in ADS is due to the capacitance variation $\Delta C$; the smaller this factor becomes, the greater the similarity between the signals. This circuit behaves as a spatiotemporal structure, where the variation of the capacity allows changing the signal in a continuous way.

%------------------------------
\subsection{Time-Modulated Varactor as $C(t)$}
Once the ideal circuit has been developed, the next step is to replace the previously employed time-varying capacitor with a varactor diode. A varactor diode provides the required variable capacitance by adjusting its reverse-bias voltage, allowing it to emulate the behavior of the ideal time-varying capacitor. This substitution enables a practical implementation of the proposed design using commercially available components. The selected component for this project will be Skyworks' SMV1236 \cite{skyworks_solutions_inc_smv123x_2020}.

\begin{figure}[!t]
    \centering
    \includegraphics[width=0.98\linewidth]{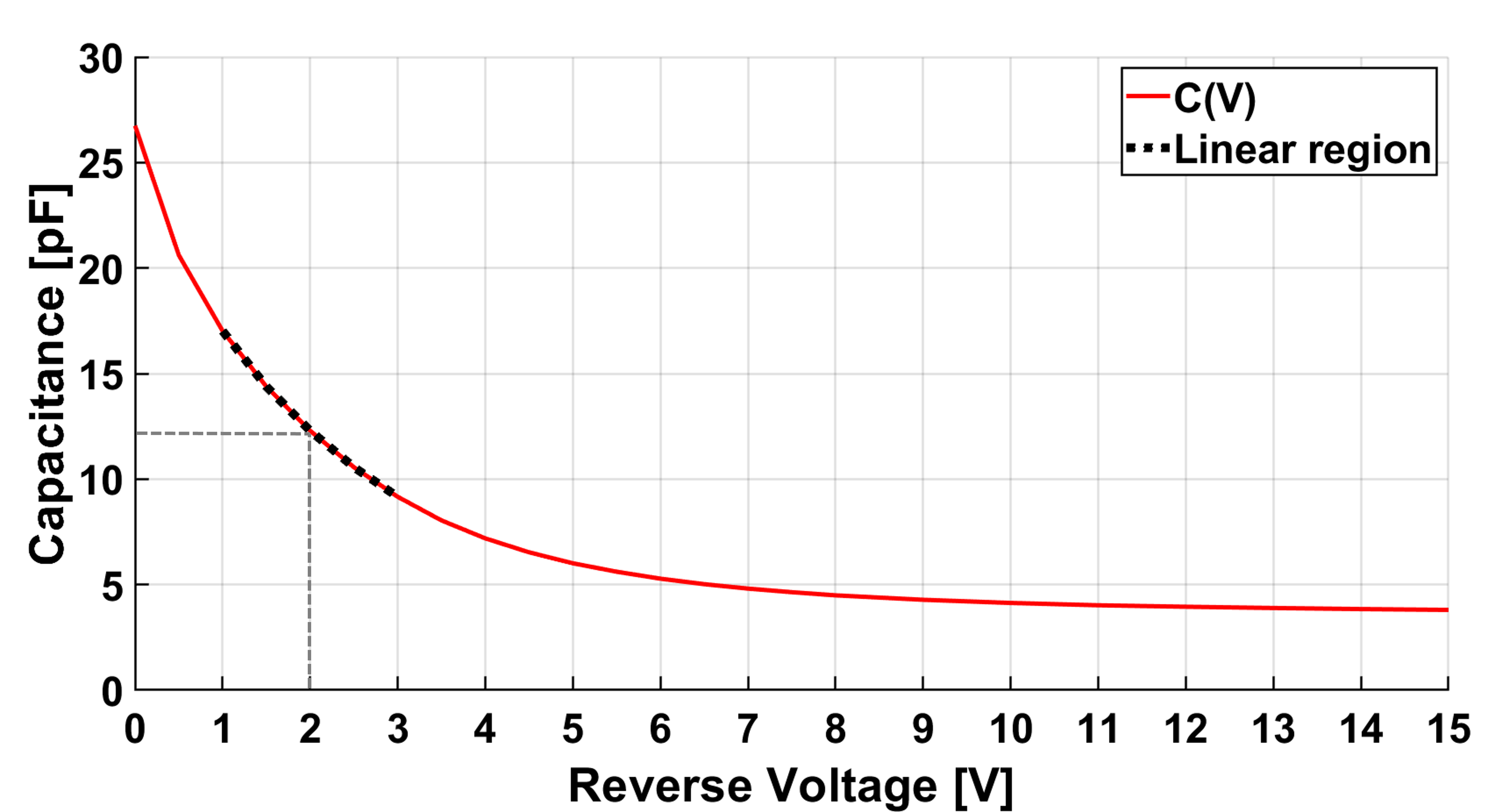}
    \caption{$C(V)$ curve of the varactor diode SMV1236 \cite{skyworks_solutions_inc_smv123x_2020}. Within the linear region, a DC voltage of $2\,\text{V}$ corresponds to a capacitance of $12.29\,\text{pF}$.}
    \label{fig:curva_varactor}
\end{figure}

Firstly, the capacitance value used should be in the linear region of the diode so that the design works correctly. In order to choose the capacitance, the $C(V)$ curve was extracted (Fig.~\ref{fig:curva_varactor}). This representation is based on the information given by the provider \cite{skyworks_solutions_inc_smv123x_2020}. As the operating bias point, we select a DC voltage of $V_{DC} = 2\,\text{V}$, which corresponds to a nominal capacitance $C_0 = 12.29\,\text{pF}$. A small-signal AC voltage is then superimposed on this DC bias, producing the desired time-varying capacitance $C(t)$ while ensuring that the varactor operates around its nominal operating point.

Several biasing networks were investigated in an attempt to incorporate the varactor diode into the original circuit topology. However, due to the location of the varactor just in the feedback loop and the simultaneous presence of the carrier and baseband excitation signals, it was not possible to establish the required DC bias without introducing undesirable coupling from the operational amplifier output. As a result, the original topology was deemed impractical for implementation, and a new circuit architecture was developed.

%%%%%%%%%%%%%%%%%%%%%%%%%%%%%%%%%%%%%%%%%%%%%%%%
\section{Alternative FM/AM Modulator Topology}
%%%%%%%%%%%%%%%%%%%%%%%%%%%%%%%%%%%%%%%%%%%%%%%%
\subsection{Ideal Circuit with a Time-Varying Capacitor}
\begin{figure}[!t]
    \centering
    \includegraphics[width=1\linewidth]{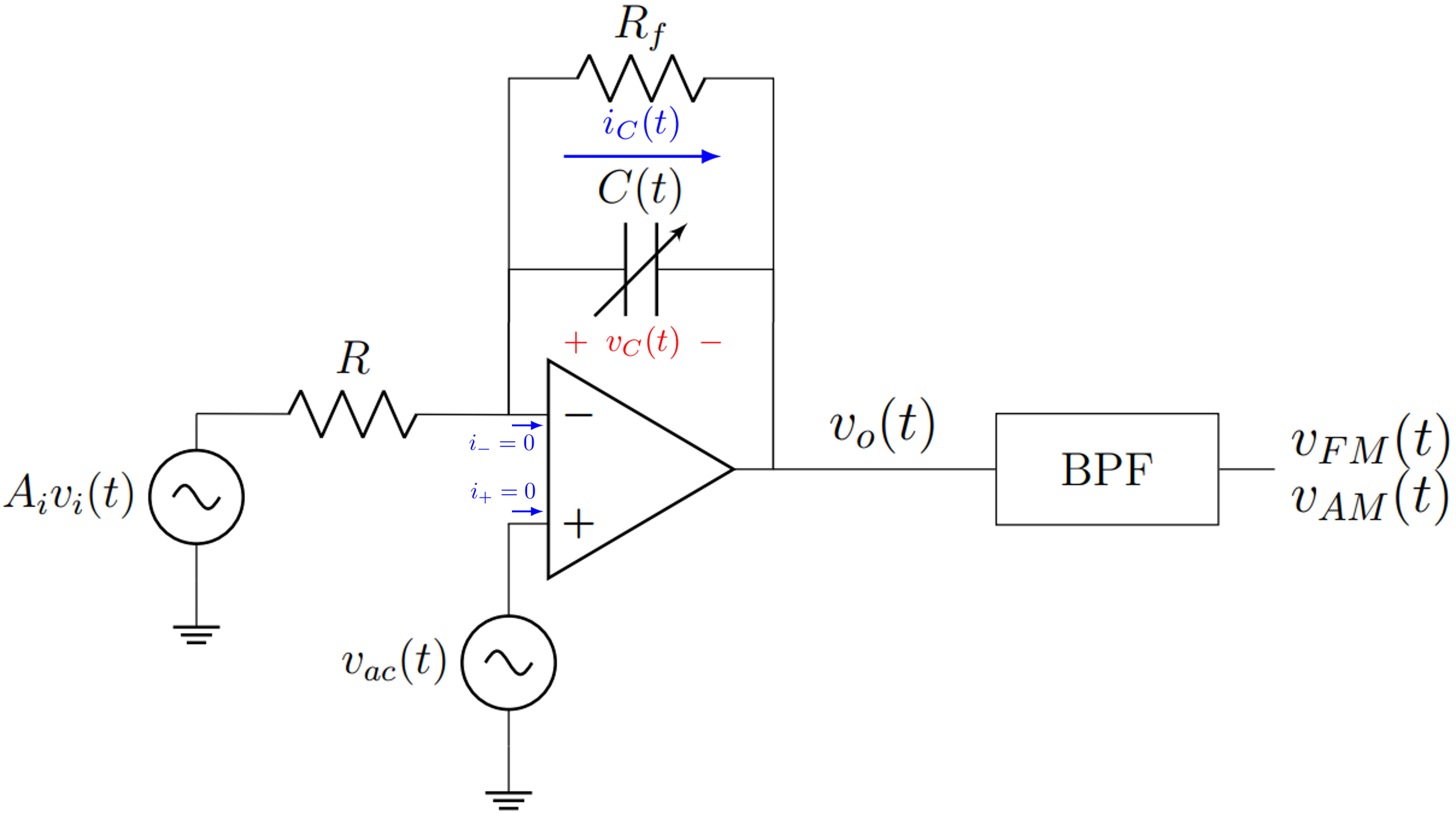}
    \caption{Second FM circuit with a time-varying capacitor.}
    \label{fig:circuito_ideal_2}
\end{figure}

As an alternative to the previous circuit, a new topology is discussed. The new topology is presented in Fig.~\ref{fig:circuito_ideal_2}. It can be regarded as a modified version of the original FM modulator shown in Fig.~\ref{fig:circuito_ideal_1}, in which the time-varying excitation is applied to the non-inverting  ``$+$" pin in the operational amplifier. This rearrangement enables the implementation of the required biasing network for the varactor diode while preserving the desired modulation mechanism.

An additional high-impedance feedback resistor $R_f$ is included to overcome the limitations of this configuration at low frequencies. Near DC, the impedance of the capacitor becomes very large, effectively opening the feedback path and preventing the proper operation of the amplifier. The resistor $R_f$ provides a DC feedback path, ensuring correct biasing and stable operation without significantly affecting the circuit behavior at the frequencies of interest due to its high impedance. The inclusion of such a resistor is a well-established design practice in analog electronics, known as a finite DC feedback integrator \cite[Ch.~2]{sedra_microelectronic_2016}.

In this way, a FM waveform can be created using the circuit in Fig.~\ref{fig:circuito_ideal_2} where, if the input baseband signal is $A_iv_i(t)$ and negligible current is assumed to flow through the high-impedance feedback resistor $R_f$, the circuit analysis reveals
\begin{equation}
    \frac{A_iv_i(t)-v_{ac}(t)}{R} = \frac{d}{dt}\left[(v_{ac}(t)-v_o(t))
    (C_0+\Delta C\cos(\omega_ct))\right],
\end{equation}
where solving for the output voltage, $v_o(t)$, one obtains
\begin{multline}
    v_o(t) = v_{ac}(t) \\
    + \frac{1}{R[C_0+\Delta C\cos(\omega_ct)]}\int^t_0(v_{ac}(\tau)-A_iv_i(\tau))d\tau.
\end{multline}

In the small-amplitude modulation regime, $\Delta C \ll C_0$, the output voltage can be simplified as
\begin{multline}
    v_o(t) \approx v_{ac}(t)\\
    +\frac{C_0-\Delta C\cos(\omega_ct)}{RC_0^2} \int^t_0(v_{ac}(\tau)-A_iv_i(\tau))d\tau.
\end{multline}

Finally, expanding all the terms and imposing $v_{ac}(t)~=~V_{ac}\cos(\omega_ct)$, the resulting voltage is
\begin{multline} \label{eq:vo_2}
    v_o(t) \approx V_{ac}\cos(\omega_ct) -\frac{A_i}{RC_0}\int^t_0v_i(\tau)d\tau -\frac{\Delta C V_{ac}}{2\omega_cRC_0^2}\sin(2\omega_ct) \\
    +\frac{V_{ac}}{\omega_cRC_0}\sin(\omega_ct) +\frac{\Delta C A_i}{RC_0^2}\cos(\omega_ct) \int^t_0v_i(\tau)d\tau.
\end{multline}

When the output signal $v_0(t)$ is bandpass-filtered, in a similar fashion to what was carried out for the first circuit, the resulting expression is
\begin{multline} \label{eq:vfm_2_completa}
    v_{FM}(t) = \text{ BPF}\{v_o(t)\} = V_{ac}\cos(\omega_ct) \\
    +\frac{V_{ac}}{\omega_cRC_0}\sin(\omega_ct) +\frac{\Delta CA_i}{RC_0^2}\cos(\omega_ct) \int^t_0v_i(\tau)d\tau.
\end{multline}
As observed, Eq.~\eqref{eq:vfm_2_completa} contains an additional term that is absent from the FM waveform obtained with the original circuit [Eq.~\eqref{eq:vfm_1}]. As a result, the bandpass-filtered output of the topology in Fig.~\ref{fig:circuito_ideal_2} is not, in general, an exact FM waveform. 

Nevertheless, the contribution of this additional term can be suppressed through an appropriate choice of the circuit parameters. In particular, by selecting the circuit parameters such that $1/(\omega_c R C_0) \gg 1$, the influence of this term is significantly reduced. Under these conditions,
\begin{equation} \label{eq:vfm_2}
    v_{FM}(t) \approx \frac{V_{ac}}{\omega_cRC_0}\sin(\omega_ct) +\frac{\Delta CA_i}{RC_0^2}\cos(\omega_ct) \int^t_0v_i(\tau)d\tau.
\end{equation}

Comparing \eqref{eq:sfm} and \eqref{eq:vfm_2}, the FM amplitude and sensitivity parameters are
\begin{equation}
    A_{FM} = \frac{V_{ac}}{\omega_cRC_0}, \quad k_{FM} = \frac{\Delta CA_i}{RC_0^2A_{FM}}
\end{equation}

However, in the regime where $1\gg 1/(\omega_c R C_0)$, the second right-hand term in Eq.~\eqref{eq:vfm_2_completa} becomes negligible compared to the first one. Consequently, it can be neglected, leading to a bandpass-filtered signal of the form
\begin{equation}
    v_{AM}(t) \approx \cos(\omega_ct) \Big[ V_{ac} + \frac{\Delta CA_i}{RC_0^2} \int^t_0v_i(\tau)d\tau\Big].
\end{equation}
The resulting waveform is an amplitude-modulated (AM) signal where the message signal $v_i$ is integrated. More specifically, the circuit in Fig.~\ref{fig:circuito_ideal_2} would act as an \emph{integrating AM modulator} provided that $1\gg 1/(\omega_c R C_0)$, instead of a \emph{FM modulator} in the regime $1\ll 1/(\omega_c R C_0)$. In the integrating AM modulator, the circuit combines temporal integration with frequency upconversion to the carrier frequency with a single circuit topology. By contrast, a conventional inverting integrator based on a static capacitor $C_0$ produces only the baseband signal
$v(t) = -1/(RC_0) \int^t_0v_i(\tau)d\tau $, requiring an additional modulation stage to generate an equivalent RF waveform.  Consequently, the circuit in Fig.~\ref{fig:circuito_ideal_2} admits two modulation schemes at once, FM and AM, which can be simply configured through the values of $\omega_c$, $R$, and $C_0$.

%%%%%%%%%%%%%%%%%%%%%%%%%%%%%%%%%%%%%%%%%%%%%%%%
\subsection{Real Circuit with a Varactor Diode}

\begin{figure}[!t]
    \centering
    \includegraphics[width=1\linewidth]{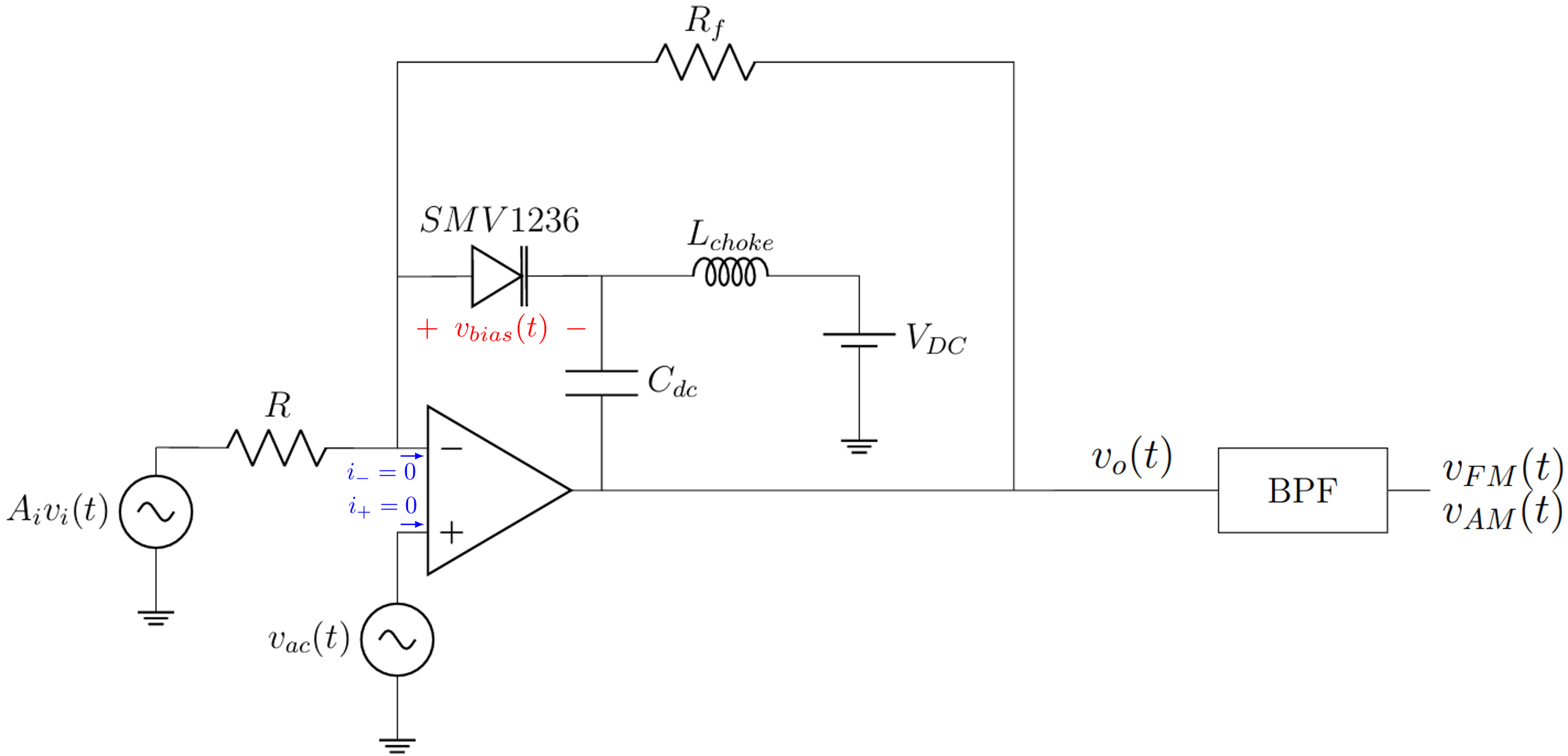}
    \caption{Complete FM circuit with a varactor diode and its bias network.}
    \label{fig:circuito_completo}
\end{figure}

For the correct implementation of the FM modulator, the time-varying capacitor $C(t)$ must be replaced by a properly biased varactor diode. Thus, the biasing network is designed so that the reverse bias voltage of the varactor diode matches the behavior of the capacitance $C(t)$ \eqref{eq:ct}. The circuit with the bias network is shown in Fig.~\ref{fig:circuito_completo}. Accordingly, the reverse bias voltage should be
\begin{equation} \label{eq:vbias}
    v_{bias}(t) = V_{DC} + V_{ac}\cos(\omega_ct),
\end{equation}
where 
\begin{equation} \label{eq:Vac}
    V_{ac} = \frac{\Delta C}{C_0}V_{DC}.
\end{equation}

In this circuit, the voltage $V_{ac}$ is already applied at the non-inverting  input ``$+$" of the operational amplifier. The virtual short-circuit condition translates the voltage $v_{ac}$ to the inverting input ``$-$" of the operational amplifier. Then, it is only necessary to include a DC voltage source at the cathode of the SMV1236 varactor to fully bias the diode. Therefore, if the polarity of this DC source is reversed, the bias voltage obtained is the one described in \eqref{eq:vbias}.

The bias network is composed of a choke inductor, $L_{choke}$, that ensures that the AC and DC paths remain isolated by imposing a high-impedance path to the RF and AC signals. On the other hand, there is a decoupling capacitor, $C_{dc}$, designed to block the DC component, thereby isolating it from the output voltage of the operational amplifier. 

\begin{figure}[!t]
  \centering
  \subfigure[]{
    \includegraphics[width=1\linewidth]{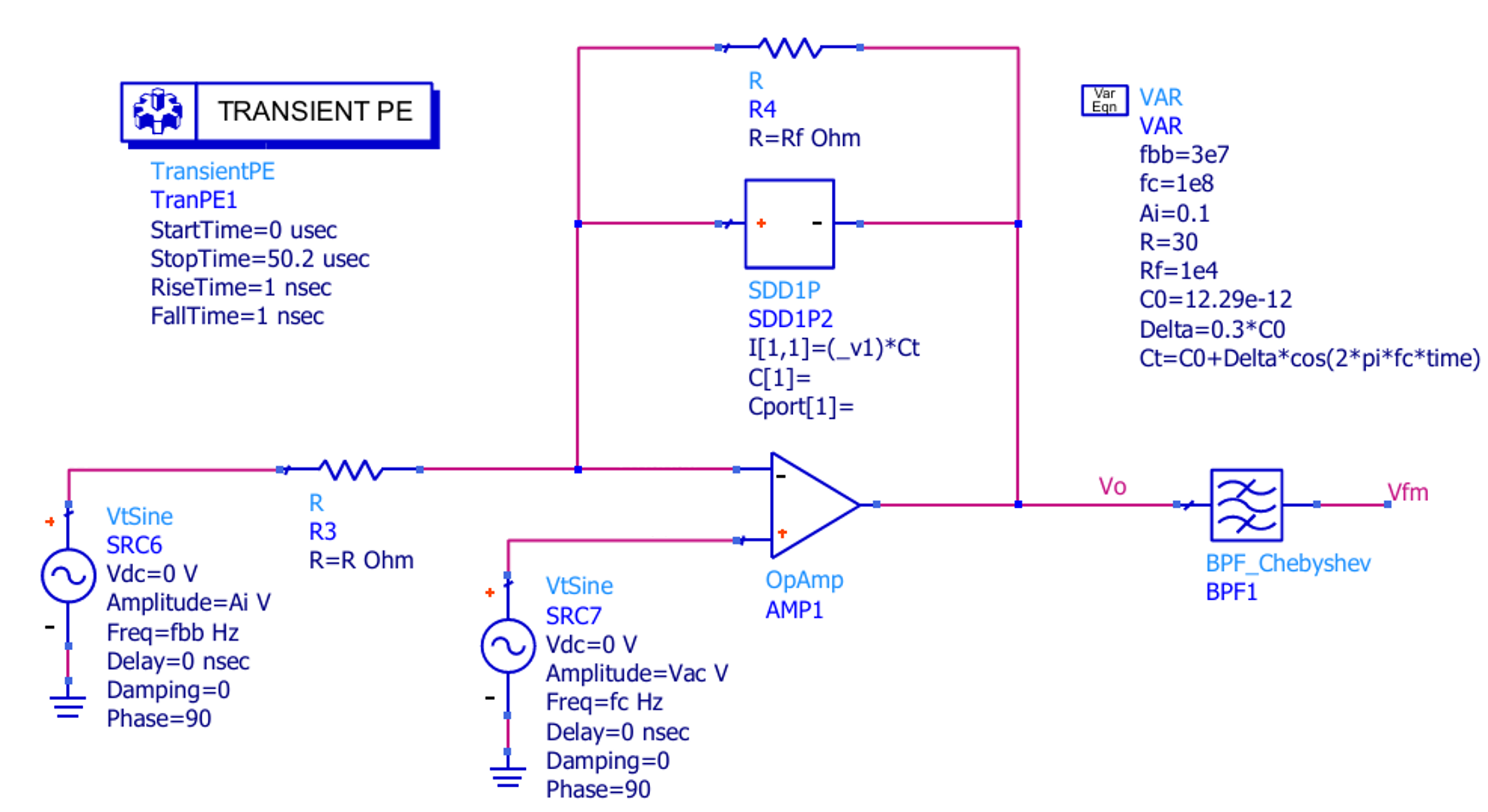}
    \label{fig:circuito_fm_ADS_2}
  }
  
  \subfigure[]{
    \includegraphics[width=1\linewidth]{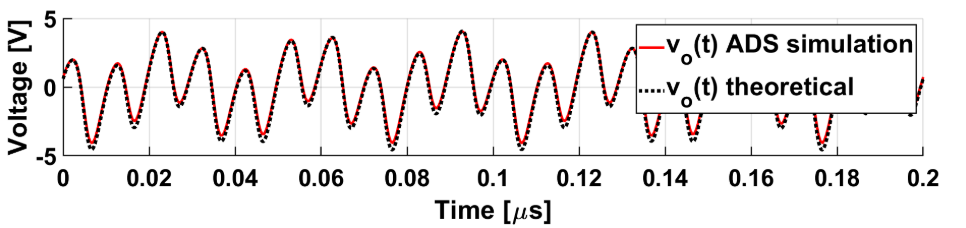}
    \label{fig:respuesta_vo_2}
  }

    \subfigure[]{
    \includegraphics[width=1\linewidth]{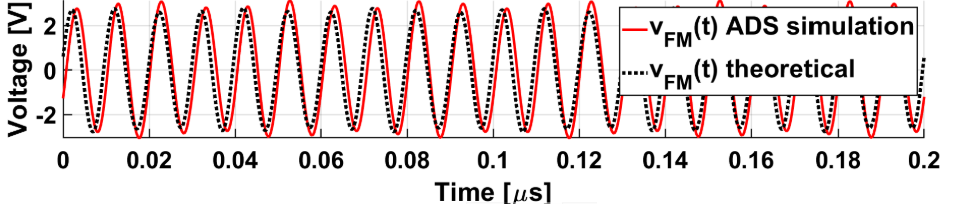}
    \label{fig:respuesta_vfm_2}
  }

  \subfigure[]{
    \includegraphics[width=1\linewidth]{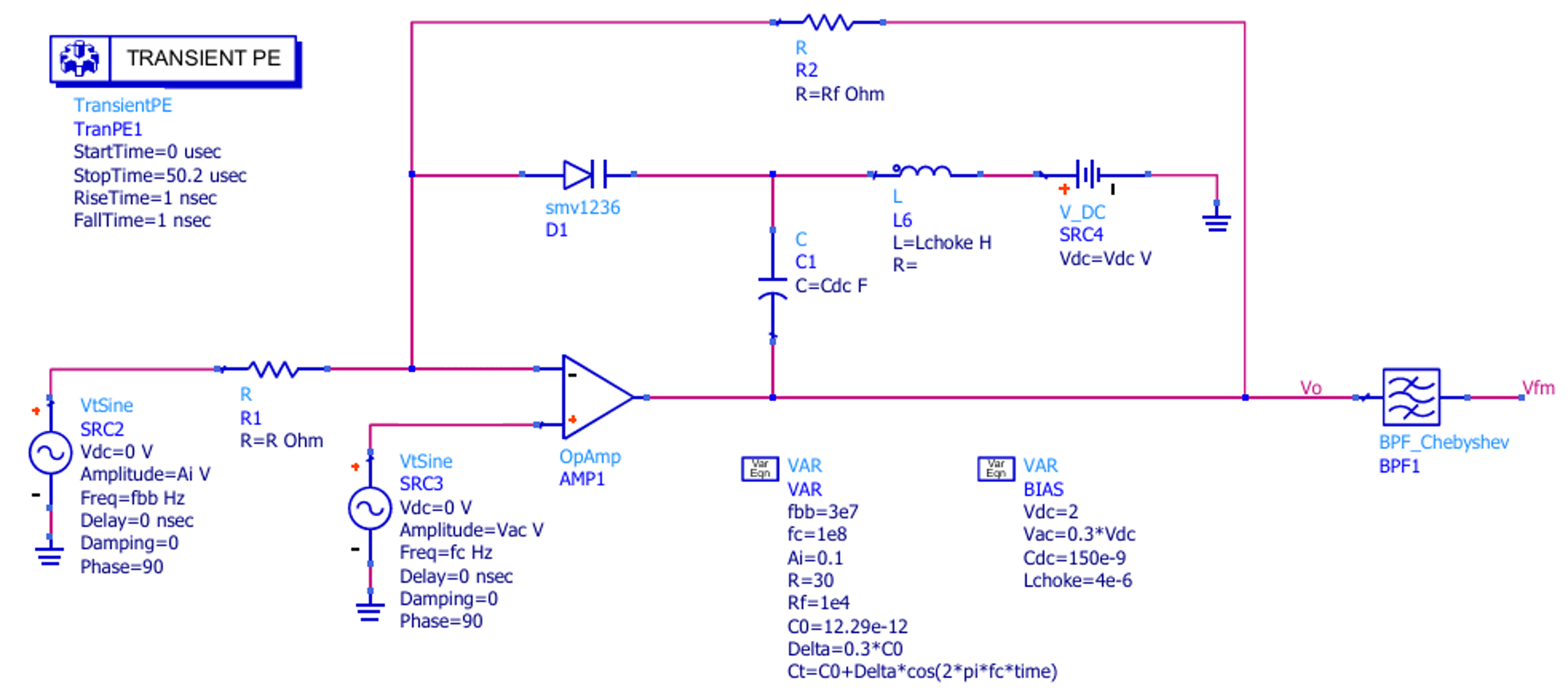}
    \label{fig:circuito_completo_ADS}
  }
  
  \subfigure[]{
    \includegraphics[width=1\linewidth]{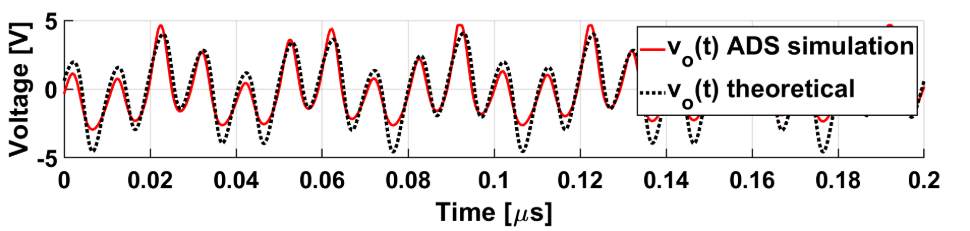}
    \label{fig:respuesta_vo}
  }

    \subfigure[]{
    \includegraphics[width=1\linewidth]{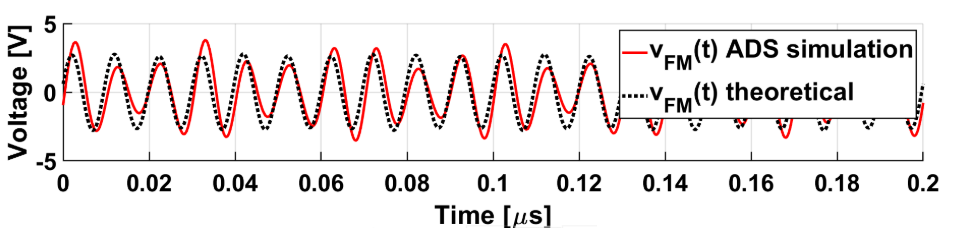}
    \label{fig:respuesta_vfm}
  }
  
  \caption{Simulated results of the circuit acting as an FM modulator. (a) Schematic of the ideal circuit in ADS. (b) $v_0$ in the ideal circuit. (c)  $v_{FM}$ in the ideal circuit. (d) Schematic of the real circuit, including its bias network. (e)  $v_0$ in the real circuit. (f)  $v_{FM}$ in the real circuit.}
  \label{fig:fm_2}
\end{figure}

\begin{figure}[!t]
  \centering
  \subfigure[]{
    \includegraphics[width=1\linewidth]{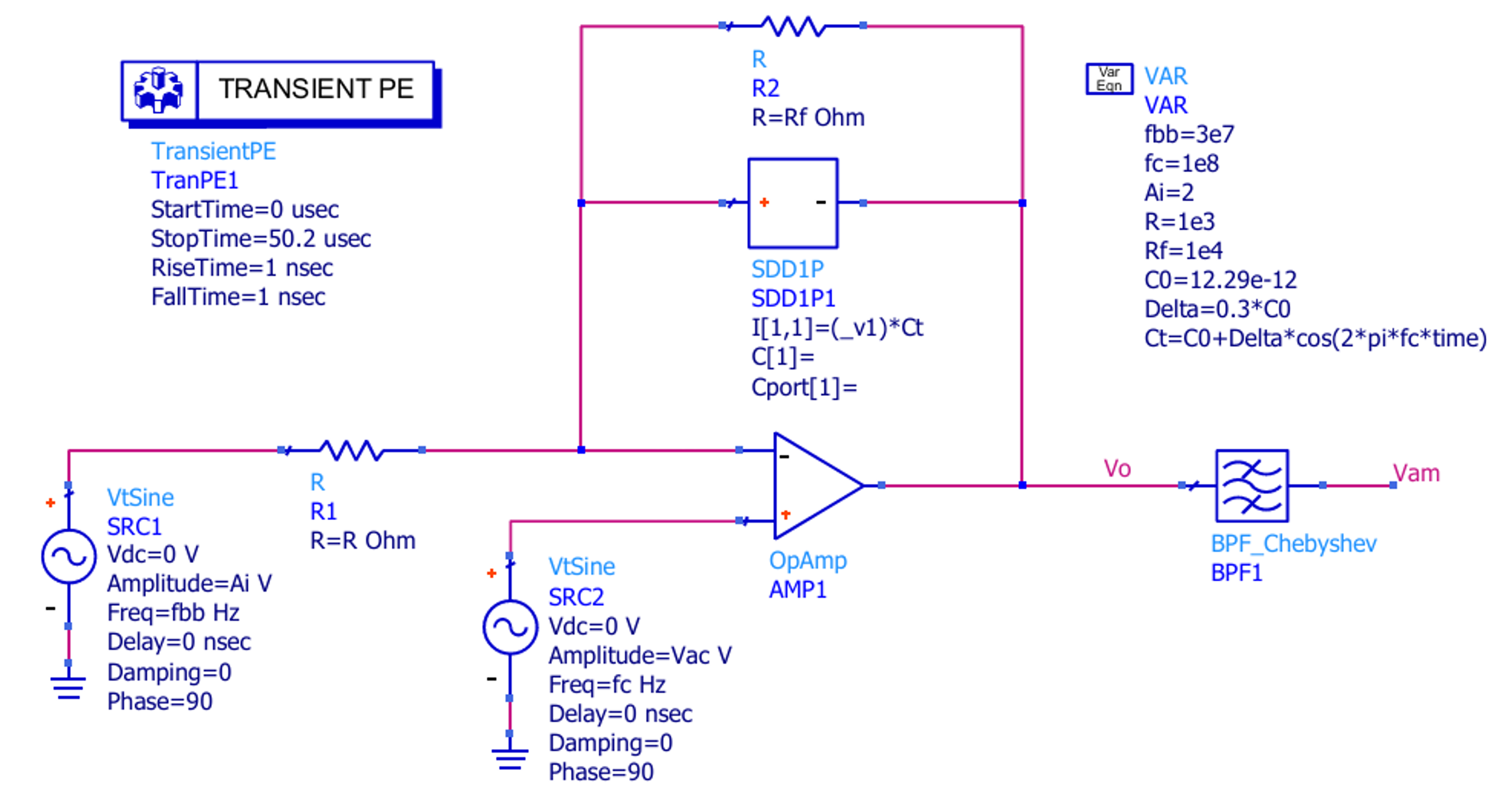}
    \label{fig:circuito_fm_ADS_2}
  }
  
  \subfigure[]{
    \includegraphics[width=0.96\linewidth]{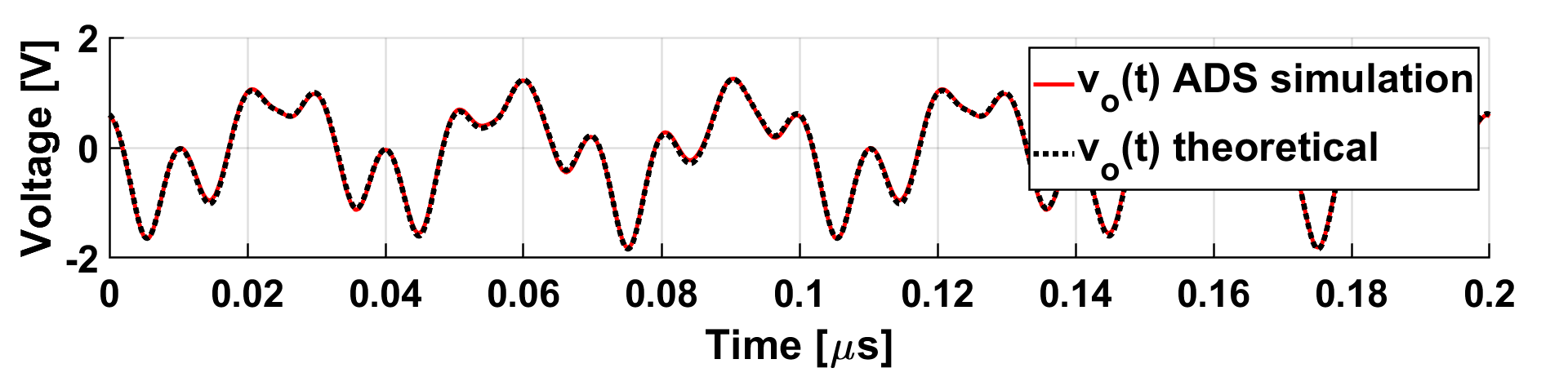}
    \label{fig:respuesta_vo_2}
  }

    \subfigure[]{
    \includegraphics[width=1\linewidth]{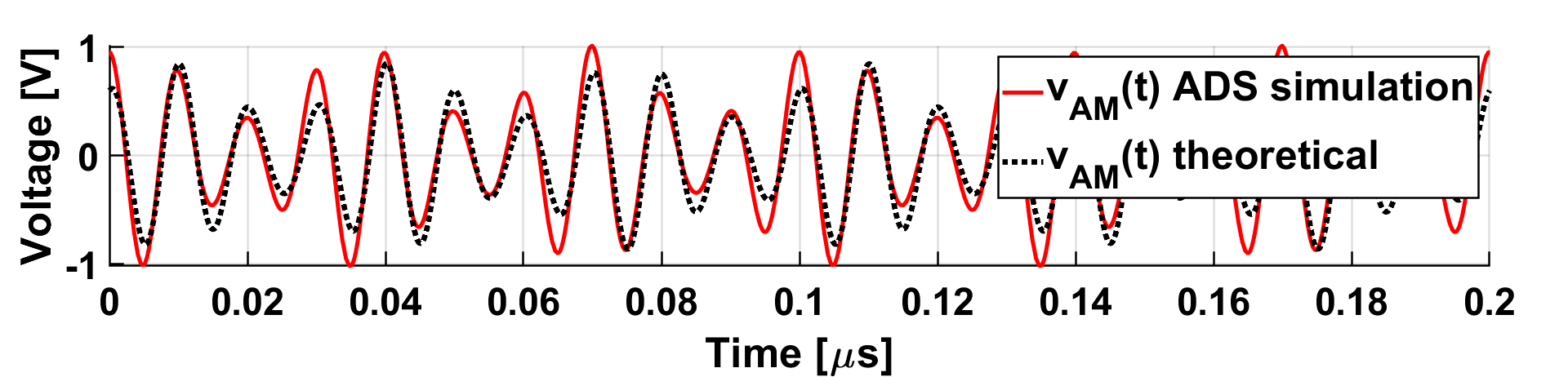}
    \label{fig:respuesta_vfm_2}
  }

  \subfigure[]{
    \includegraphics[width=1\linewidth]{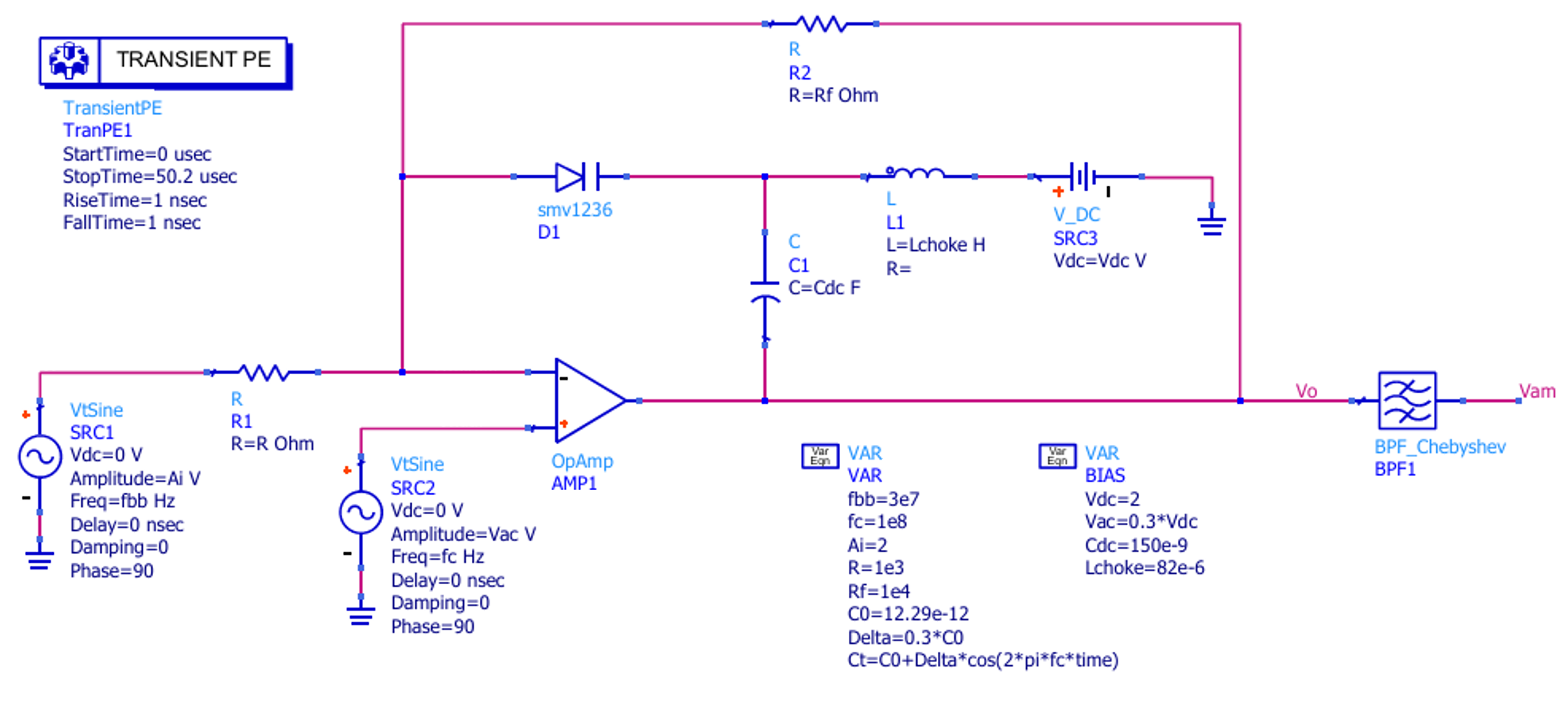}
    \label{fig:circuito_completo_ADS}
  }
  
  \subfigure[]{
    \includegraphics[width=1\linewidth]{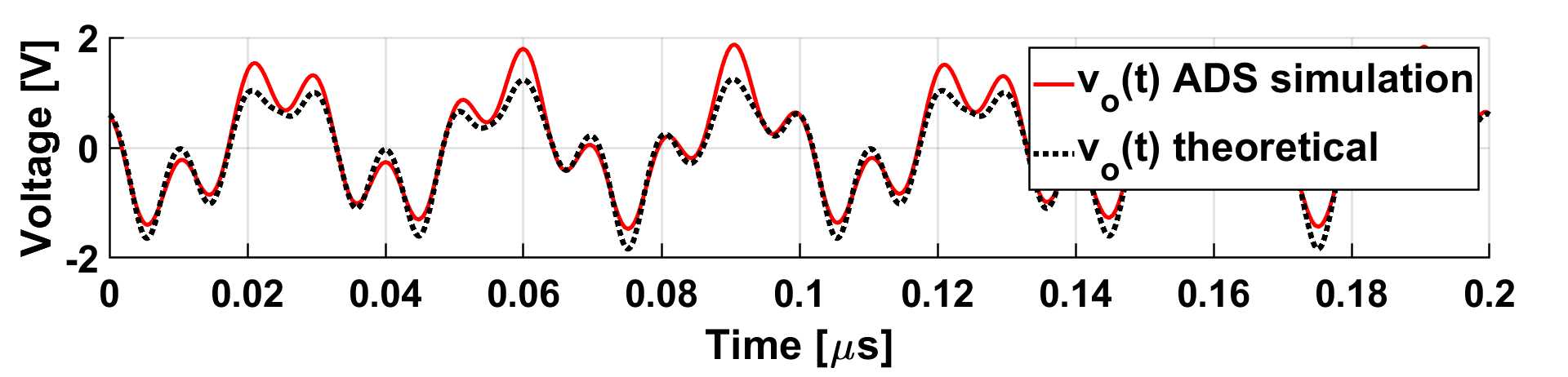}
    \label{fig:respuesta_vo}
  }

    \subfigure[]{
    \includegraphics[width=1\linewidth]{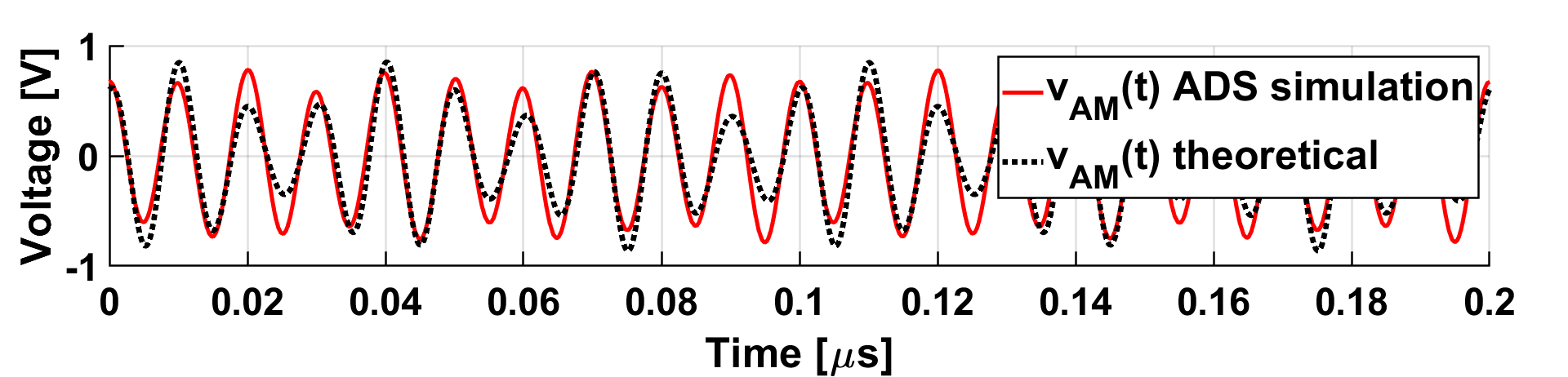}
  }
  
  \caption{Simulated results of the circuit acting in the integrating-AM operation mode. (a) Schematic of the ideal circuit in ADS. (b) $v_0$ in the ideal circuit. (c)  $v_{AM}$ in the ideal circuit. (d) Schematic of the real circuit, including its bias network. (e) $v_0$ in the real circuit. (f)  $v_{AM}$ in the real circuit.}
  \label{fig:am_2}
\end{figure}
%%%%%%%%%%%%%%%%%%%%%%%%%%%%%%%%%%%%%%%%%%%%%%%%
\subsection{Simulation Results}

Since this second version of the FM modulator is implementable, the correct operation of the alternative circuit topology in Figs.~\ref{fig:circuito_ideal_2} and \ref{fig:circuito_completo} is evaluated. Both the FM and integrating-AM modulation modes are tested in simulation. In both scenarios, we consider the following common circuit parameters: $C_0 = 12.29\,\text{pF}$,  $k_{FM} = 0.3\omega_c$, carrier frequency $f_c = 100\,\text{MHz}$, feedback resistor $R_f = 10$ k$\Omega$,  and decoupling capacitor $C_{dc}=150\,\text{nF}$. A modulation factor of 30\% is applied, corresponding to $\Delta C = 0.3C_0$ and $V_{ac} = 0.6$ V. The message (baseband) signal is $v_i = A_i \cos(2\pi f_{bb}t)$, with $f_{bb} = 30$ MHz.

The obtained simulation results for the circuit in Fig.~\ref{fig:circuito_ideal_2} operating as an FM modulator are presented in Fig.~\ref{fig:fm_2}. In the FM case, the resistor is set to $R = 30\, \Omega$, the input voltage is $A_i = 0.1$ V, and the choke inductor is $L_{choke} = 4\, \mu$H. Given that the capacitance is $C_0=12.29\,\text{pF}$ this results in $1/(\omega_c R C_0) = 4.32$,  which can be considered significantly higher than $1$ at all practical effects. As before, the block ``BPF\_Chebyshev'' includes the bandpass filter implemented in Fig.~\ref{fig:bpf}(a). 

Figs.~\ref{fig:fm_2}(a)-(c) show the schematic and simulation results of the ideal FM modulator circuit (without the bias network). Figs.~\ref{fig:fm_2}(e)-(f) show the corresponding results for the practical real circuit (with the bias network). In both the ideal and real circuits, the FM waveform $v_{FM}(t)$ is obtained after bandpass filtering $v_0(t)$. As shown in the time-domain waveforms, the instantaneous frequency of the bandpass-filtered signal varies continuously with the modulating signal, producing the expected compression and expansion of the waveform cycles that characterize frequency modulation. Moreover, the good agreement between the theoretical predictions (black dashed curves) and the simulation results (red curves) further validates the proposed circuit topology as an alternative implementation of an FM modulator. Although replacing the ideal $C(t)$ in Figs.~\ref{fig:fm_2}(a)-(c) with the real time-modulated varactor in Figs.~\ref{fig:fm_2}(d)-(f) slightly degrades the resulting signals, which is a product of the imperfect isolation between the DC and RF signals, the overall FM waveform is well preserved.

Subsequently, we test the integrating-AM operation mode in Fig.~\ref{fig:am_2}. To enter this operation mode, we set $R = 1\,\text{k}\Omega$, accompanied by $A_i = 2$ V and $L_{choke} = 82\, \mu$H. Knowing that $C_0=12.29\,\text{pF}$, it yields $1/(\omega_c R C_0) = 0.13 \ll 1$. Figs.~\ref{fig:am_2}(a)-(c) illustrate the schematic and simulation results of the ideal circuit (without the bias network), whereas Figs.~\ref{fig:am_2}(e)-(f) likewise show the corresponding results for the real circuit (with the bias network). Unlike the FM operation mode, the instantaneous frequency of the bandpass-filtered waveform in  the AM-integrating mode remains essentially constant, while its envelope follows the integrated modulating signal (a sinusoid of 30 MHz in this case). The close agreement between the ideal [Figs.~\ref{fig:am_2}(b)-(c)] and practical real [Figs.~\ref{fig:am_2}(e)-(f)]  implementations confirms that the addition of the bias network has only a minor impact on the integrating-AM response, preserving the expected envelope evolution and thus the correct operation of the time-modulated varactor diode.

%%%%%%%%%%%%%%%%%%%%%%%%%%%%%%%%%%%%%%%%%%%%%%%%
\section{Experimental Results}
\begin{figure}[!t]
  \centering
  \subfigure[]{
    \includegraphics[width=1\linewidth]{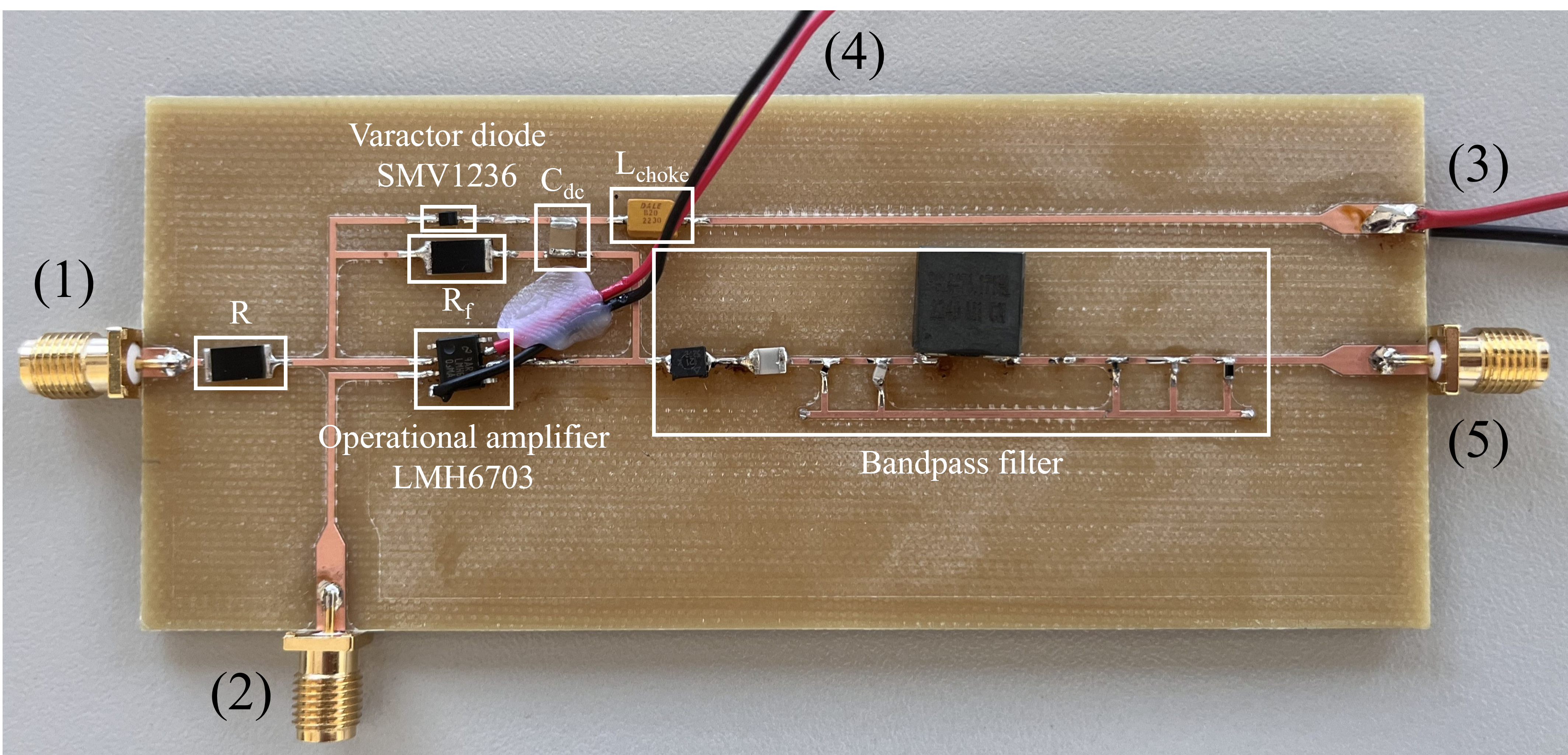}
    \label{fig:placa_soldada}
  }
  \subfigure[]{
    \includegraphics[width=0.4\linewidth]{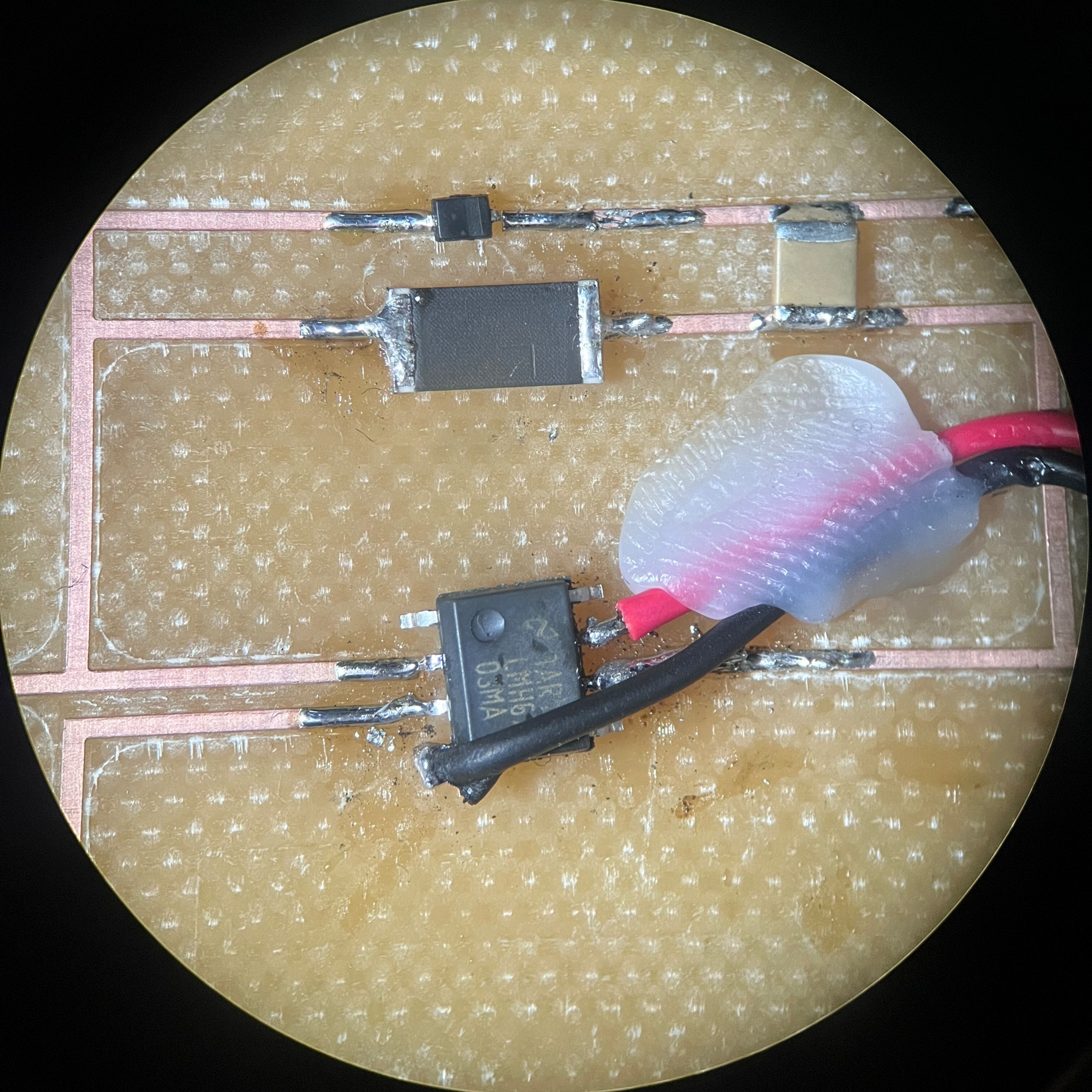}
    \label{fig:placa_integrador}
  }%
    \subfigure[]{
    \includegraphics[width=0.4\linewidth]{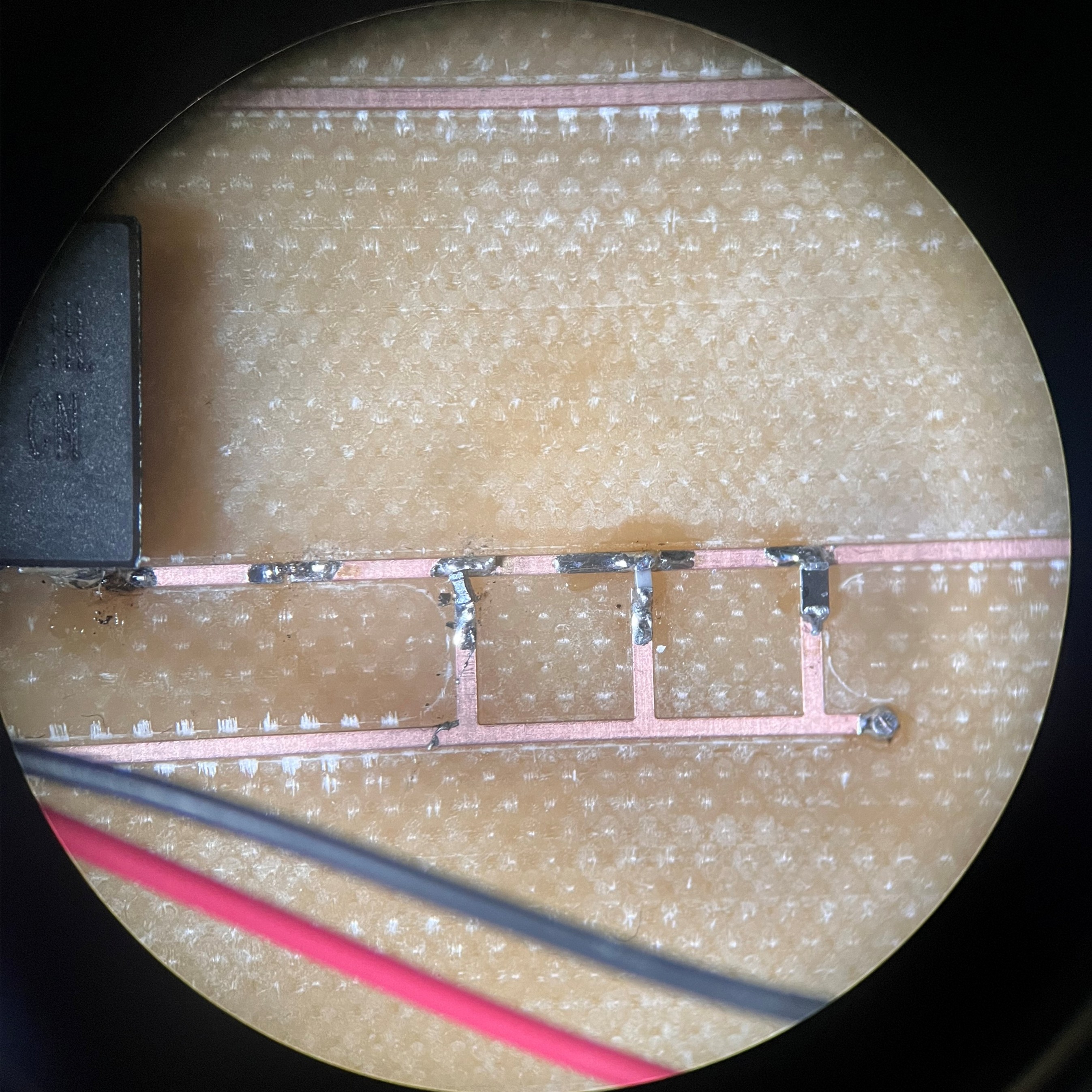}
    \label{fig:placa_filtro}
  }%
  \caption{Prototype of the dual-mode FM/AM modulator implemented on a PCB using microstrip technology. (a) Assembled PCB, with five ports: (1) input baseband signal $v_i(t)$, (2) input AC carrier $v_{ac}(t)$, (3) DC input $V_{DC}$ feeding the varactor, (4) DC input feed for the operational amplifier, and (5) output $v_{FM}(t)$ or $v_{AM}(t)$ waveforms. (b) Microscopic view of the integrator area, showing the varactor diode and operational amplifier. (c) Microscopic view of the bandpass filter area.}
  \label{fig:placa}
\end{figure}

%%%%%%%%%%%%%%%%%%%%%%%%%%%%%%%%%%%%%%%%%%%%%%%%
After verifying in simulation the correct operation of the dual-mode FM/AM modulator, we proceed with its fabrication. Figure~\ref{fig:placa} shows the manufactured prototype, along with close-up views of selected areas, highlighting key components such as the operational amplifier, the varactor diode, and the bandpass filter. The printed circuit board (PCB), fabricated and then measured in the facilities of the RFCAS research group at Universidad Autónoma de Madrid, is implemented using microstrip technology on an FR-4 substrate. The design employs a reference impedance of 50~$\Omega$, corresponding to a microstrip trace width of 2.69~mm. This width is significantly large for the components size, so a progressive width transition (taper) has been made. Therefore, the trace width is locally reduced to 0.6~mm facilitate the soldering process. In this way, the input and output ports of the board are compatible with the SMA connectors of 50~$\Omega$, while the area accommodating the components features a narrower width.

The circuit was assembled manually, focusing on the soldering of the active devices: the SMV1236 varactor diode \cite{skyworks_solutions_inc_smv123x_2020} and the LMH6703 operational amplifier \cite{texas_instruments_lmh6703_2016}. Circuit connectivity was achieved using three SMA connectors for the input signals and the FM output. Additionally, different connection wires were integrated for the DC input, the operational amplifier power supply, and the ground (GND) reference.

\begin{figure}[!t]
  \centering
  \subfigure[]{
    \includegraphics[width=0.485\linewidth]{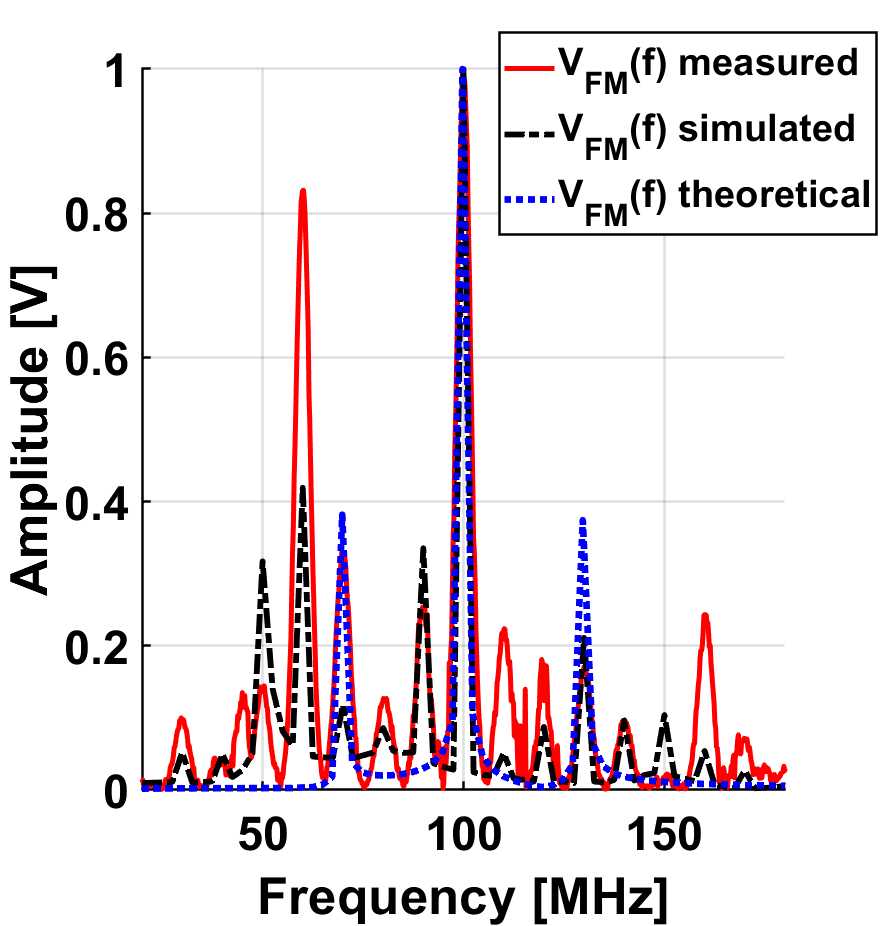}
    \label{fig:medida_fm}
  }%
    \subfigure[]{
    \includegraphics[width=0.485\linewidth]{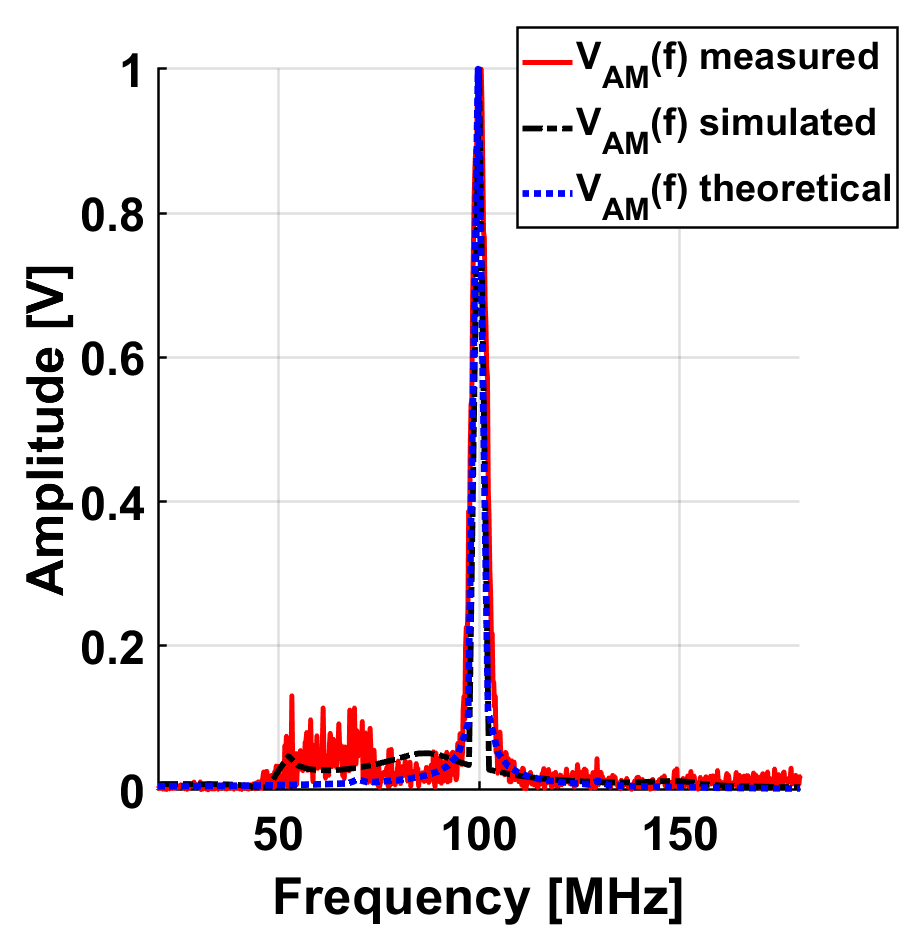}
    \label{fig:medida_am}
  }%
  \caption{Measured spectrum of the prototype in Fig.~\ref{fig:placa} for the (a) FM and (b) integrating-AM operation modes. Theory, simulations and measurements are compared.}
  \label{fig:medidas}
\end{figure}

Following the circuit assembly, the laboratory measurements were conducted. The experimental setup to measure the prototype in Fig.~\ref{fig:placa} utilized two RF signal generators (Aim-TTi TGR2050 and Agilent N9310A) for the input signals, baseband ``(1)"  and carrier ``(2)", two DC power supplies (Promax FA-376 and FAC-662B) for the DC biasing of the varactor ``(3)" and operational amplifier ``(4)", and a spectrum analyzer Rohde \& Schwarz FSL ``(5)" to capture the  spectral response of the  generated output waveforms.

The DC levels for the biasing of the  LMH6703 operational amplifier \cite{texas_instruments_lmh6703_2016} and SMV1236 varactor diode \cite{skyworks_solutions_inc_smv123x_2020} are $\pm5\,\text{V}$ and $2\,\text{V}$, respectively. Moreover, it should be pointed out that one of the RF generators used had a limitation in its amplitude range. It could not reach the $0.6\,\text{V}$ of the AC input voltage amplitude, $v_{ac}(t)$, so it has to be readjusted to $0.5\,\text{V}$. Following the established relation in \eqref{eq:Vac}, this adjustment implies a modification in the capacitance modulation $V_{ac}= V_{DC}\, \Delta C/C_0 = 2\,\text{V}\cdot0.25C_0/C_0=0.5\,\text{V}$.
Consequently, the capacitance modulation factor decreased from $30\%$ to $25\%$.

Subsequently, we proceed to perform the measurements. Since an oscilloscope with sufficient bandwidth (200 MHz) was not available to capture the time-domain response, the experimental validation was carried out in the frequency domain. Consequently, the experimental results shown in Fig.~\ref{fig:medidas} correspond to the output spectrum measured with a spectrum analyzer and are compared with the theoretical and simulated results. For a consistent comparison, the Fast Fourier Transform (FFT) was applied to the theoretical and simulated time-domain signals to obtain their corresponding frequency spectra.

Figure~\ref{fig:medidas}(a) shows the normalized results for the circuit operating as an FM modulator ($R = 50\,\Omega$ and $A_i = 0.5\,\text{V}$), while Fig.~\ref{fig:medidas}(b) presents the corresponding results in integrating-AM operation mode ($R = 10\,\text{k}\Omega$ and $A_i =1\,\text{V}$). As seen, the agreement between theory (blue curves), simulations in ADS of the circuit with the bias network (black curves), and measurements (red curves) is very good in the integrating-AM mode [Fig.~\ref{fig:medidas}(b)]. This operation mode produces a ``simpler" spectrum, with fewer harmonics than the FM mode. In the FM mode [Fig.~\ref{fig:medidas}(a)], the agreement is particularly good for the dominant harmonic contributions ($70, 100, 130\,\text{MHz}$). Additional spectral components, spaced by $10\,\text{MHz}$, are observed in the measurements. These components are not predicted by the approximate theoretical model, which retains only the three dominant harmonics, but they are captured by the circuit simulations. Their presence is mainly attributed to the nonlinear behavior of the practical circuit components, particularly the varactor diode and the operational amplifier.

The experimental results further confirm the correct operation of the dual-mode FM/AM modulator, demonstrating the feasibility of implementing frequency modulation through a time-varying capacitance scheme combined with an inverting integrator. These results highlight the potential of time-modulation techniques as a promising alternative design paradigm for realizing modulation functionalities, opening new opportunities for communication and radar systems.

%%%%%%%%%%%%%%%%%%%%%%%%%%%%%%%%%%%%%%%%%%%%%%%%
\section{\label{sec:Conc} Conclusion}
%%%%%%%%%%%%%%%%%%%%%%%%%%%%%%%%%%%%%%%%%%%%%%%%
This paper details the analysis, design, and experimental realization of a frequency and amplitude modulator, operating at a center frequency of $100\,\text{MHz}$, that exploits novel temporal modulation schemes. The architecture of the circuit is based on the time-varying capacitance of a varactor diode, combined with an operational amplifier in an inverting-integrator topology and a bandpass filter. Analytical expressions describing the FM waveform generated by the proposed modulator have been derived. Moreover, two closely related circuit topologies were analyzed and compared, the second extending the functionality of the first by supporting both FM and integrating-AM operation modes. Finally, a prototype of the proposed FM modulator was fabricated on a PCB in microstrip technology and its performance was evaluated. The agreement between theory, simulations, and measurements validates the proposed circuit topology and confirms the feasibility of implementing frequency modulation using a time-varying capacitance scheme. Beyond demonstrating a new modulator architecture, this work expands the application of time-modulated circuits beyond nonreciprocal filters and isolators, opening new opportunities for the development of modern high-performance architectures for future communication and radar systems.

\ifCLASSOPTIONcaptionsoff
  \newpage
\fi

%%%%%%%%%%%%%%%%%%%%%%%%%%%%%%%%%%%%%%%
%\begin{thebibliography}{1}
\bibliographystyle{IEEEtran}
\bibliography{IEEEabrv, references}

\end{document}